%% file: quant-cat-notes.tex
\documentclass[12pt]{amsproc}
\pdfoutput=1

\DeclareMathOperator{\Lie}{Lie}

\input{preamble.tex}

\newcommand{\emp}{\textsl}
\def\vocab#1{{\textsl{#1}}%
\index{#1}}

\def\upR{\xygraph{
!{0;/r1.0em/:}
[u(0.5)]
!{\xoverv=<}
}}

\def\upL{\xygraph{
!{0;/r1.0em/:}
[u(0.5)]
!{\xunderv=<}
}}

\def\upS{\xygraph{
!{0;/r1.0em/:}
[u(0.5)]
!{\huncross=>}
}}

\def\unknot{\smash{\!\xygraph{
!{0;/r1.0em/:}
!{\vcap-}
!{\vcap=>}
}}}

\def\hopflink{
\! \xygraph{
!{0;/r0.5em/:}
[u(1)]
!{\vover}
!{\vover-}
[uur]!{\hcap[2]=<}
[l]!{\hcap[-2]=>}
} \! }

\newcounter{needrefs}

\newcounter{looks}

\makeindex

\begin{document}

\title{Lectures on Knot Homology and Quantum Curves}

\author[S. Gukov]{Sergei Gukov}
\address{California Institute of Technology}
\email{gukov@theory.caltech.edu}

\author[I. Saberi]{Ingmar Saberi}
\email{isaberi@caltech.edu}

\thanks{This survey was presented in a series of lectures first at University of Notre Dame (2012 summer school on \textsl{Topology and Field Theories} at the Center for Mathematics), then at Stanford (summer school on \textsl{Holomorphic curves and low-dimensional topology}), and in other versions in Hamburg, Oberwolfach, Lisbon, and Heidelberg.
}

\begin{abstract}
Besides offering a friendly introduction to knot homologies and quantum curves,
the goal of these lectures is to review some of the concrete predictions that follow from
the physical interpretation of knot homologies.
In particular, it allows one to answer questions like
\textsl{Is there a direct relation between Khovanov homology and the $A$-polynomial of a knot?}
which would not have been asked otherwise.
We will explain that the answer to this question is ``yes'' and introduce
a certain deformation of the planar algebraic curve defined by the zero locus of the $A$-polynomial.
This novel deformation leads to a categorified version of the Generalized Volume Conjecture
that completely describes the ``color behavior'' of the colored $\lie{sl}(2)$ knot homology
and, eventually, to a similar version for the colored HOMFLY homology.
Furthermore, this deformation is strong enough to distinguish mutants,
and its most interesting properties include relation to knot contact homology and knot Floer homology.
\end{abstract}

\maketitle

\tableofcontents

\section*{Foreword}

An alternative title of these lecture notes could be ``Categorification and Quantization.''
These lectures, however, will by no means serve as a complete introduction to the two topics of \vocab{quantization} and \vocab{categorification}.  Each of these words represents not so much a single idea as a broad tool, program, or theme in physics and mathematics, and both are areas of active research and are still not fully understood.
One could easily give a full one-year course on each topic separately.

Rather, the goal of these lectures is to serve as an appetizer: to give a glimpse of the ideas behind quantization and categorification, by focusing on very concrete examples and giving a working knowledge of how these ideas are manifested in simple cases. It is our hope that the resulting discussion will remain accessible and clear while still shedding some light on these complex ideas, and that the interest of the reader will be piqued.

Imagine the category of finite-dimensional vector spaces and linear maps. To each object in this category is naturally associated a number, the dimension of that vector space. Replacing some collection of vector spaces with a collection of numbers in this way can be thought of as a \vocab{decategorification}: by remembering only the dimension of each space, we keep some information, but lose all knowledge about (for instance) morphisms between spaces. In this sense, decategorification forgets about geometry.

Categorification can be thought of as the opposite procedure. Given some piece of information (an invariant of a topological space, for instance), one asks whether it arises in some natural way as a ``decategorification'': a piece of data extracted out of a more geometrical or categorical invariant, which may carry more information and thus be a finer and more powerful tool. An answer in the affirmative to this question is a categorification of that invariant.

Perhaps the most familiar example of categorification at work can be seen in the reinterpretation of the Euler characteristic
as the alternating sum of ranks of homology groups,
\beq
\chi(M) = \sum_{k\geq 0} (-1)^k \rank H_k(M) \,.
\label{Eulerhom}
\eeq
Thus, the homology of a manifold $M$ can be seen, in a sense, as a categorification of its Euler characteristic: a more sophisticated and richly structured bearer of information, from which the Euler characteristic can be distilled in some natural way. Moreover, homology theories are a far more powerful tool than the Euler characteristic alone for the study and classification of manifolds and topological spaces.
This shows that categorification can be of practical interest: by trying to categorify invariants, we can hope to construct \emp{stronger} invariants.

While the idea of categorification is rooted in pure mathematics, it finds a natural home in the realm of
topological quantum field theory (TQFT) as will be discussed in Section~\ref{sec:categorification}.
For this, however, we first need to understand what ``quantum'' means and to explain the quantization program
that originated squarely within physics. Its basic problem is the study of the transition between classical and quantum mechanics. The classical and quantum pictures of a physical system make use of entirely different and seemingly unconnected mathematical formalisms.  In classical mechanics, the space of possible states of the system is a symplectic manifold, and observable quantities are smooth functions on this manifold. The quantum mechanical state space, on the other hand, is described by a Hilbert space~$\Hil$, and observables are elements of a noncommutative algebra of operators acting on~$\Hil$. Quantization of a system is the construction of the quantum picture of that system from a classical description, as is done in a standard quantum mechanics course for systems such as the harmonic oscillator and the hydrogen atom. Therefore, in some sense, quantization allows one to interpret quantum mechanics as ``modern symplectic geometry.'' We will give a more full introduction to this idea in Section~\ref{sec:quantization}.

One main application of the ideas of quantization and categorification is to representation theory,
where categorification, or ``geometrization,'' leads naturally to the study of geometric representation theory \cite{CGbook}.
Another area of mathematics where these programs bear much fruit is low-dimensional topology, which indeed is often called ``quantum'' topology. This is the arena in which we will study the implications of quantization and categorification, primarily for the reason that it allows for many concrete and explicit examples and computations. Specifically, almost all of our discussion will take place in the context of knot theory. The reader should not, however, be deceived into thinking of our aims as those of knot theorists! We do not discuss quantization and categorification for the sake of their applications to knot theory; rather, we discuss knot theory because it provides a window through which we can try and understand quantization and categorification.

\vspace{1cm}
\noindent\textbf{Acknowledgements} \\
We would like to thank Tudor Dimofte, Hiroyuki Fuji, Lenhard Ng,
Marko Sto$\check{\text{s}}$i$\acute{\text{c}}$, Piotr Su{\l}kowski, Cumrun Vafa, Edward Witten, Don Zagier
for enlightening discussions and enjoyable collaborations on subjects considered in these notes.
We are also grateful to Tudor Dimofte, Lenhard Ng, and Piotr Su{\l}kowski for their comments on the draft.

This survey was presented in a series of lectures first at University of Notre Dame
(2012 summer school on \textsl{Topology and Field Theories} at the Center for Mathematics),
then at Stanford (summer school on \textsl{Holomorphic curves and low-dimensional topology}),
and in other versions in Hamburg (summer school on \textsl{Strings and Fundamental Physics}),
Oberwolfach (Workshop  \textsl{Low-Dimensional Topology and Number Theory}),
Lisbon (program on \textsl{Recent Advances in Topological Quantum Field Theory}),
and Heidelberg (workshop \textsl{Mathematics and Physics of Moduli spaces}).
We would like to thank the organizers of all these meetings for their generous support, accommodations, and collaborative working environment.
This work is supported in part by DOE Grant DE-FG03-92-ER40701FG-02 and in part by NSF Grant PHY-0757647.
Opinions and conclusions expressed here are those of the authors and do not necessarily reflect the views of funding agencies.

\section{Why knot homology?}
\label{sec:intro}

\begin{figure}
\includegraphics[width=1.5in]{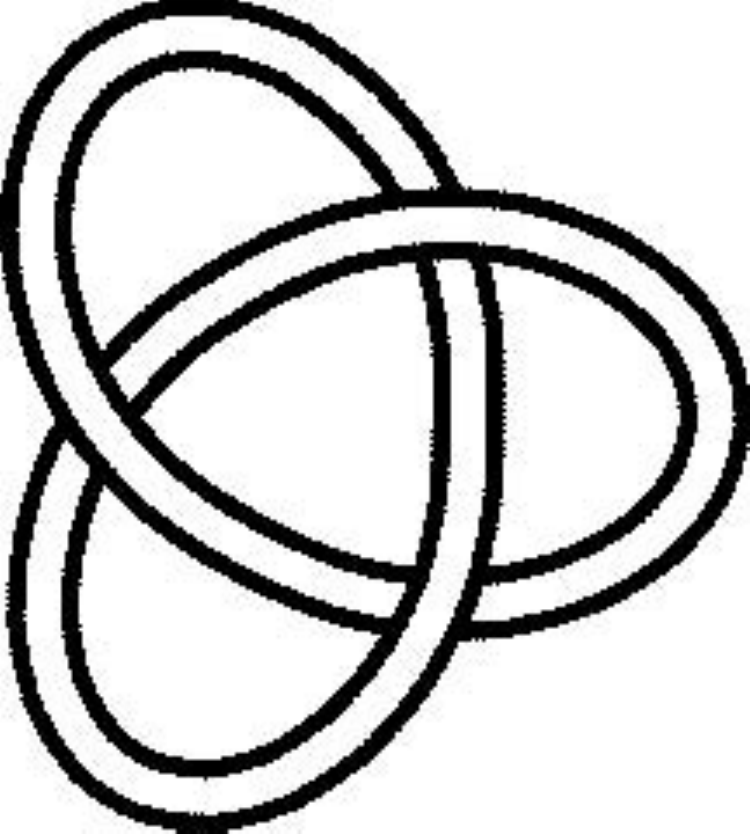}
\caption{The trefoil knot $3_1$. (KnotPlot image from {\tt http://katlas.math.toronto.edu}.)\label{trefoil}}
\end{figure}

\begin{figure}
\includegraphics[width=2in]{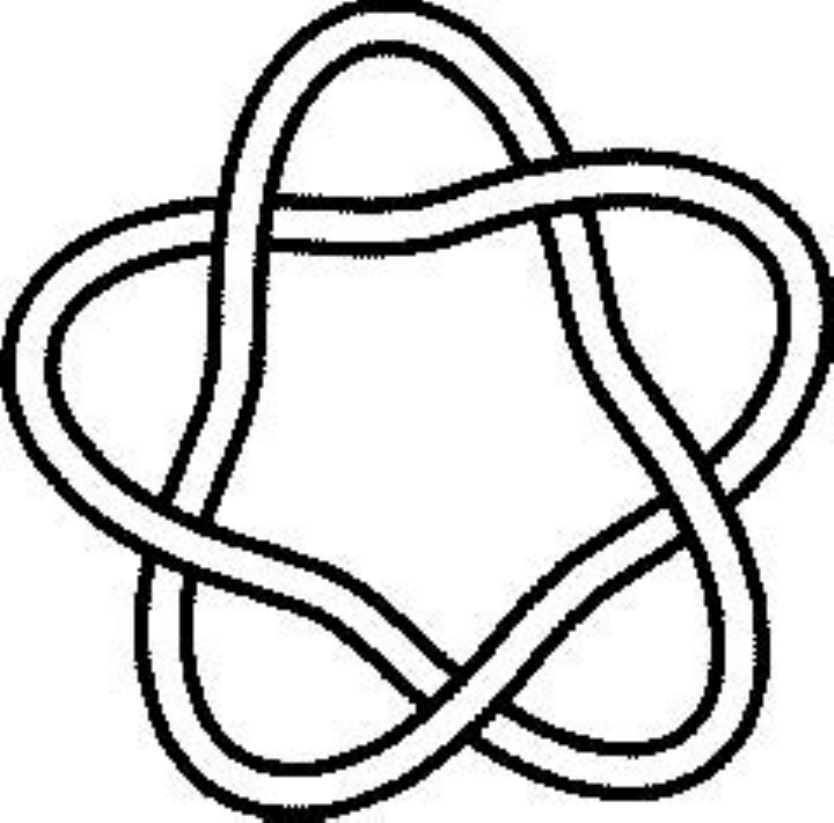}
\quad
\includegraphics[width=2in]{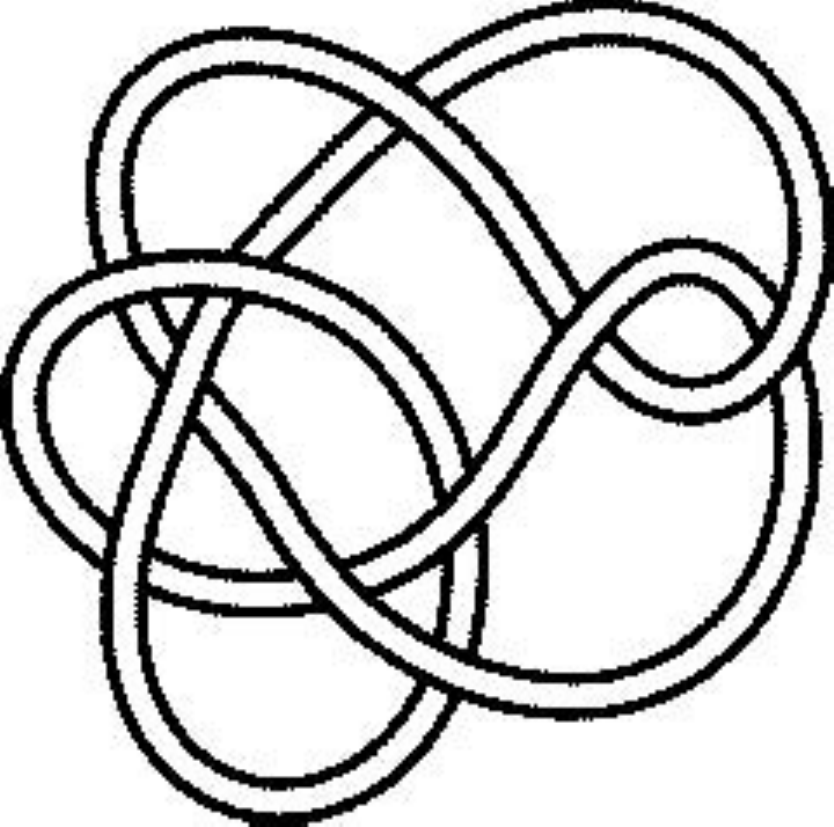}
\caption{The knots $5_1$ and $10_{132}$. (KnotPlot images from {\tt http://katlas.math.toronto.edu}.)\label{twoknots}}
\end{figure}

A \vocab{knot} is a smooth embedding of a circle $S^1$ as a submanifold of~$S^3$:
\beq
k: S^1 \inj S^3, \quad K \defeq \im k,
\eeq
see {\sl e.g.\/} Figures \ref{trefoil} and \ref{twoknots} for some simple examples.
Likewise, a \vocab{link} is defined as an embedding of several copies of $S^1$.

In attempting to classify knots, one of the most basic tools is a \vocab{knot invariant}\textsl{:} some mathematical object that can be associated to a knot, that is always identical for equivalent knots. In this way, one can definitively say that two knots are distinct if they possess different invariants. The converse, however, is not true; certain invariants may fail to distinguish between knots that are in fact different. Therefore, the arsenal of a knot theorist should contain a good supply of different invariants. Moreover, one would like invariants to be as ``powerful'' as possible; this just means that they should capture nontrivial information about the knot. Obviously, assigning the number 0 to every knot gives an invariant, albeit an extremely poor one!

Given the goal of constructing knot invariants, it may be possible to do so most easily by including some extra structure to be used in the construction. That is, one can imagine starting not simply with a knot, but with a knot ``decorated'' with additional information: for instance, a choice of a Lie algebra
$\lie{g}=\Lie(G)$ and a representation~$R$ of~$\lie{g}$. It turns out that this additional input data from representation theory does in fact allow one to construct various invariants (numbers, vector spaces, and so on), collectively referred to as \vocab{quantum group invariants}. A large part of these lectures will consist, in essence, of a highly unorthodox introduction to these quantum group invariants.

The unorthodoxy of our approach is illustrated by the fact that we fail completely to address a natural question: what on earth do (for instance) the quantum $\lie{sl}(N)$ invariants have to do with~$\lie{sl}(N)$? Representation theory is almost entirely absent from our discussion; we opt instead to look at an alternative description of the invariants, using a concrete combinatorial definition in terms of so-called \vocab{skein relations}. A more full and traditional introduction to the subject would include much more group theory, and show the construction of the quantum group invariants in a way that makes the role of the additional input data~$\lie{g}$ and~$R$ apparent~\cite{RT,WittenCS}.
That construction involves assigning a so-called ``quantum $R$-matrix'' to each crossing in a knot diagram in some manner, and then taking a trace around the knot in the direction of its orientation. The connection to representation theory is made manifest; the resulting invariants, however, are the same.

\begin{example}
Suppose that we take an oriented knot together with the Lie algebra $\lie{g}=\lie{sl}(N)$ and its fundamental $N$-dimensional representation. With this special choice of extra data, one constructs the quantum $\lie{sl}(N)$ invariant, denoted $P_N(K;q)$. Although it makes the connection to representation theory totally obscure, one can compute~$P_N(K;q)$ directly from the knot diagram using the following skein relation:
\beq q^N P_N(\upR) - q^{-N} P_N (\upL) = (q-q^{-1}) P_N(\upS). 	\label{N-skein}\eeq
(Note that we will sometimes write $P_N(K)$ for the polynomial $P_N$ associated to  the knot or link~$K$, suppressing the variable~$q$; no confusion should arise.) For now, one can think of~$q$ as a formal variable. The subdiagrams shown in~\eqref{N-skein} should be thought of as depicting a neighborhood of one particular crossing in a planar diagram of an oriented knot; to apply the relation, one replaces the chosen crossing with each of the three shown partial diagrams, leaving the rest of the diagram unchanged.

To apply this linear relation, one also needs to fix a normalization, which can be done by specifying $P_N$ for the unknot. Here, unfortunately, several natural choices exist. For now, we will choose
\beq P_N(\unknot) = \frac{q^N-q^{-N}}{q-q^{-1}} = \underbrace{q^{-(N-1)} + q^{-(N-3)} + \cdots + q^{N-1}}_{N\text{ terms}}.
		\label{unnorm} \eeq
This choice gives the so-called \vocab{unnormalized} $\lie{sl}(N)$ polynomial.
Notice that, given any choice of $P_N(\unknot)$ with integer coefficients, the form of the skein relation implies that $P_N(q) \in \Z[q,q^{-1}]$ for every knot.

Notice further that, with the normalization~\eqref{unnorm}, we have
\beq P_N(\unknot) \xrightarrow[q\goesto 1]{} N, \eeq which is the dimension of the representation~$R$ with which we decorated the knot, the fundamental of~$\lie{sl}(N)$. We remark that this leads to a natural generalization of the notion of dimension, the so-called \vocab{quantum dimension} $\dim_q(R)$ of a representation~$R$, which arises from the quantum group invariant constructed from~$R$ evaluated on the unknot.

Equipped with the above rules, let us now try to compute $P_N(q)$ for some simple links. Consider the \vocab{Hopf link}, consisting of two interlocked circles:
\[ \xygraph{
!{0;/r1.75pc/:}
!{\vover}
!{\vover-}
[uur]!{\hcap[2]=<}
[l]!{\hcap[-2]=>}
} \]
Applying the skein relation to the upper of the two crossings, we obtain:
\beq
q^N P_N\Big[\!\!
\underset{\text{Hopf link}}%
{\xygraph{
!{0;/r0.75em/:}
[u(1)]
!{\vover}
!{\vover-}
[uur]!{\hcap[2]=<}
[l]!{\hcap[-2]=>}
}}
\!\!\Big]
 - q^{-N} P_N \Big[\!\!
\underset{\text{two unknots}}%
{\xygraph{
!{0;/r0.75em/:}
[u(1)]
!{\vunder}
!{\vover-}
[uur]!{\hcap[2]=<}
[l]!{\hcap[-2]=>}
}}
\!\!\Big]
= (q-q^{-1}) P_N \Big[\!\!
\underset{\text{one unknot}}%
{\xygraph{
!{0;/r0.75em/:}
[u(1)]
!{\xunoverv}
!{\vover-}
[uur]!{\hcap[2]=<}
[l]!{\hcap[-2]=>}
}}
\!\!\Big].
\label{hopf1}
\eeq
This illustrates a general feature of the skein relation, which occurs for knots as well as links: In applying the relation to break down any knot diagram into simpler diagrams, one will in fact generally need to evaluate $P_N$ for links rather than just for knots, since application of the relation~\eqref{N-skein} may produce links with more than one component. This means that the normalization~\eqref{unnorm} is not quite sufficient; we will need to specify~$P_N$ on $k$ unlinked copies of the unknot, for $k\geq 1$.

As such, the last of our combinatorial rules for computing $P_N(q)$ concerns its behavior under disjoint union:
\beq
P_N(\unknot\sqcup K) = P_N(\unknot)\cdot P_N(K),
\label{disjoint}
\eeq
where $K$ is any knot or link. Here, the disjoint union should be such that~$K$ and the additional unknot are not linked with one another.
\paragraph{\emp{Caution:}} The discerning reader will notice that our final rule~\eqref{disjoint} is not linear, while the others are, and so is not respected under rescaling of~$P_N(q)$. Therefore, if a different choice of normalization is made, it will \emp{not} remain true that
$ P_N(k\text{ unknots}) = [P_N(\unknot)]^k$. The nice behavior~\eqref{disjoint} is particular to our choice of normalization~\eqref{unnorm}. This can be expressed by saying that, in making a different normalization, one must remember to normalize only one copy of the unknot.

To complete the calculation we began above, let's specialize to the case~$N=2$. Then we have
\beq
P_2(\unknot) = q^{-1} + q \qquad\implies P_2(\!
\xygraph{
!{0;/r0.5em/:}
[u(1)]
!{\vunder}
!{\vover-}
[uur]!{\hcap[2]=<}
[l]!{\hcap[-2]=>}
}
\!)
= (q^{-1} + q)^2 = q^{-2} + 2 + q^2.
\eeq
Applying the skein relation~\eqref{hopf1} then gives
\beq
\begin{aligned} q^2 P_2(\hopflink) &= q^{-2} (q^{-2} + 2 + q^2) + (q-q^{-1})(q+q^{-1}) \\
&= q^{-4} + q^{-2} + 1 + q^2,
\end{aligned}
\eeq
so that
\beq
P_2(\hopflink) = q^{-6} + q^{-4} + q^{-2} + 1.
\eeq
We are now ready to compute the $\lie{sl}(N)$ invariant for any link.
\end{example}

From the form of the rules that define this invariant, it is apparent that dependence on the parameter~$N$ enters the knot polynomial only by way of the combination of variables~$q^N$. As such, we can define the new variable $a\defeq q^N$, in terms of which our defining relations become
\beq a P_{a,q}(\upR) - a^{-1} P_{a,q} (\upL) = (q-q^{-1}) P_{a,q}(\upS), 	\label{HOMFLY-skein}\eeq
\beq P_{a,q}(\unknot) = \frac{a-a^{-1}}{q-q^{-1}}.		\label{HOMFLY-unnorm} \eeq
Together with the disjoint union property, these rules associate to each oriented link~$K$ a new invariant~$P_{a,q}(K)$ in the variables~$a$ and~$q$, called the (unnormalized) \vocab{HOMFLY-PT polynomial} of the link \cite{HOMFLY}. This is something of a misnomer, since with the normalization~\eqref{HOMFLY-unnorm} the HOMFLY-PT invariant will in general be a rational expression rather than a polynomial. We have traded the two variables $q$, $N$ for~$q$ and~$a$.

For various special choices of the variables~$a$ and~$q$, the HOMFLY-PT polynomial reduces to other familiar polynomial knot invariants:

\begin{itemize}

\item $a=q^N$, of course, returns the quantum $\lie{sl}(N)$ invariant $P_N(q)$.

\item With the particular choice $a=q^2$ ($N=2$), the HOMFLY-PT polynomial becomes the classical \vocab{Jones polynomial} $J (L;q) \equiv P_2 (q)$,
\beq
J (K;q) = P_{a=q^2,q} (K).
\label{JonesPq2}
\eeq
Discovered in 1984~\cite{jones}, the Jones polynomial is one of the best-known polynomial knot invariants, and can be regarded as the ``father'' of quantum group invariants; it is associated to the Lie algebra~$\lie{sl}(2)$ and its fundamental two-dimensional representation.

\item $a=1$ returns the \vocab{Alexander polynomial} $\Delta (K;q)$, another classical knot invariant. This shows that the HOMFLY-PT polynomial generalizes the $\lie{sl}(N)$ invariant, in some way: the evaluation $a=1$ makes sense, even though taking $N=0$ is somewhat obscure from the standpoint of representation theory.

\end{itemize}

Now, the attentive reader will point out a problem: if we try and compute the Alexander polynomial, we immediately run into the problem that~\eqref{HOMFLY-unnorm} requires $P_{1,q}(\unknot)=0$. The invariant thus appears to be zero for every link! However, this does not mean that the Alexander polynomial is trivial. Remember that, since the skein relations are linear, we have the freedom to rescale invariants by any multiplicative constant. We have simply made a choice that corresponds, for the particular value $a=1$, to multiplying everything by zero.

This motivates the introduction of another convention: the so-called \vocab{normalized} HOMFLY-PT polynomial is defined by performing a rescaling such that
\beq
P_{a,q}(\unknot) = 1. 	\label{HOMFLY-norm}
\eeq
This choice is natural on topological grounds, since it associates 1 to the unknot independent of how the additional input data, or ``decoration,'' is chosen. (By contrast, the \vocab{unnormalized} HOMFLY-PT polynomial assigns the value~$1$ to the empty knot diagram.) Taking $a=1$ in the {\sl normalized\/} HOMFLY-PT polynomial returns a nontrivial invariant, the Alexander polynomial.

\begin{exercise}	\label{hw:HOMFLY}
Compute the normalized and unnormalized HOMFLY-PT polynomials for the trefoil knot~$K=3_1$ (Fig.~\ref{trefoil}). Note that one of these will actually turn out to be polynomial!

Having done this, specialize to the case $a=q^2$ to obtain the normalized and unnormalized Jones polynomials for the trefoil. Then specialize to the case $a=q$. Something nice should occur! Identify what happens and explain why this is the case.
\begin{proof}[Solution]
Applying the skein relation for the HOMFLY-PT polynomial to one crossing of the trefoil knot gives
\[
a P_{a,q}(3_1) - a^{-1}  P_{a,q}(\unknot) = (q-q^{-1})  P_{a,q}(\hopflink).
\]
Then,  applying the relation again to the Hopf link (as in the above example) gives
\[
a P_{a,q}(\hopflink) - a^{-1} P_{a,q}
(\!
\xygraph{
!{0;/r0.5em/:}
[u(1)]
!{\vunder}
!{\vover-}
[uur]!{\hcap[2]=<}
[l]!{\hcap[-2]=>}
}
\!)
= (q-q^{-1})  P_{a,q}(\unknot).
\]
Therefore, for the unnormalized HOMFLY-PT polynomial,
\[
 P(3_1) = a^{-2}  P(\unknot) + a^{-2} (q-q^{-1}) \left[ a^{-1}  P(\unknot)^2  + (q-q^{-1})  P(\unknot)\right].
\]
which becomes
\[
P(3_1) = \frac{a-a^{-1}}{q-q^{-1}} \left[  a^{-2} q^2+ a^{-2} q^{-2} - a^{-4}  \right].
\]
The normalized HOMFLY-PT polynomial is simply the quantity in brackets. Specializing to $a=q^2$ gives the unnormalized Jones polynomial:
\beq
P_2(3_1) = \frac{q^2-q^{-2}}{q-q^{-1}} \left[ q^{-2} + q^{-6} - q^{-8} \right].
\label{trefJones}
\eeq
Again, the normalized Jones polynomial is the factor in square brackets. Finally, we specialize to $a=q$, obtaining $P=1$ in both the normalized and unnormalized cases! This is connected to the fact that $a=q$ corresponds to constructing the $\lie{sl}(1)$ invariant, which must be vacuous since the Lie algebra is trivial.
\end{proof}
\end{exercise}

\begin{remark}
The study of this subject is made more difficult by the preponderance of various conventions in the literature. In particular, there is no agreement at all about standard usage with regard to the variables for polynomial invariants. Given ample forewarning, this should not cause too much confusion, but the reader must always be aware of the problem. In particular, it is extremely common for papers to differ from our conventions by the replacement
\beq
a\mapsto a^{1/2},\ q\mapsto q^{1/2},
\label{aqconf1}
\eeq
halving all powers that occur in knot polynomials. Some authors also make the change
\beq
a\mapsto a^{-1},\ q\mapsto q^{-1},
\label{aqconf2}
\eeq
and some make both.
\end{remark}

We have by now seen a rich supply of knot polynomials, which can be straightforwardly computed by hand for simple enough diagrams, and are easy to write down and compare. One might then ask about the value of attempting to categorify at all. Given such simple and powerful invariants, why would one bother trying to replace them with much more complicated ones?

\begin{figure}
\includegraphics[height=0.95in]{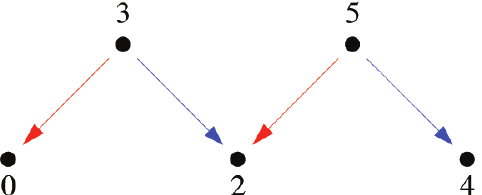}
\qquad
\includegraphics[height=0.95in]{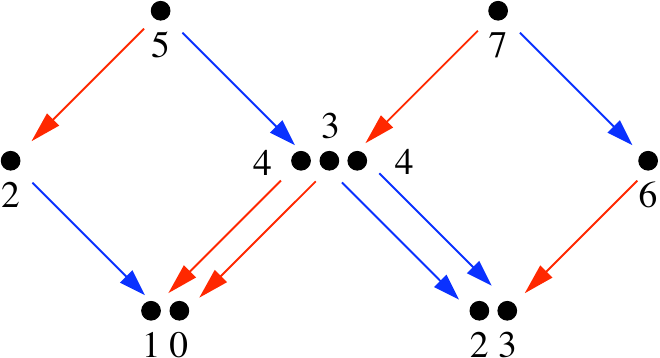}
\caption{The HOMFLY homology for knots $5_1$ and $10_{132}$. (Images from \cite{DGR}.)\label{twoknotsHOMFLY}}
\end{figure}

The simple answer is that the HOMFLY-PT polynomial and its relatives, while powerful, are not fully adequate for the job of classifying all knots up to ambient isotopy. Consider the two knot diagrams shown in Fig.~\ref{twoknots}, which represent the knots~$5_1$ and~$10_{132}$ in the Rolfsen classification.
While the knots are not equivalent, they have identical Alexander and Jones polynomials! In fact, we have
\beq
P_{a,q}(5_1) = P_{a,q}(10_{132}) \,.
\label{PP5110132}
\eeq
and, therefore, all specializations --- including all $\lie{sl}(N)$ invariants --- will be identical for these two knots.
Thus, even the HOMFLY-PT polynomial is not a perfect invariant and fails to distinguish between these two knots. This motivates us to search for a finer invariant. Categorification, as we shall see, provides one. Specifically, even though Jones, Alexander and HOMFLY-PT polynomials in this particular example fail to distinguish $5_1$ and~$10_{132}$ knots, their respective categorifications do (cf. Figure~\ref{twoknotsHOMFLY}).\\

Before we step into the categorification era, let us make one more desperate attempt to gain power through polynomial knot invariants.
To this end, let us introduce not one, but a whole sequence of knot polynomials $J_n(K; q) \in \Z[q,q^{-1}]$
called the \vocab{colored Jones polynomials}. For each non-negative integer $n$, the $n$-colored Jones polynomial of a knot~$K$ is the quantum group invariant associated to the decoration $\lie{g}=\lie{sl}(2)$ with its $n$-dimensional representation~$V_n$. $J_2(K;q)$ is just the ordinary Jones polynomial. In \vocab{Chern-Simons theory} with gauge group $G=\SU(2)$, we can think of $J_n(K;q)$ as the expectation value of a Wilson loop operator on~$K$, colored by the $n$-dimensional representation of~$\SU(2)$ \cite{WittenCS}.

Moreover, the colored Jones polynomial obeys the following relations, known as \vocab{cabling formulas}, which follow directly from the rules of Chern-Simons TQFT:
\beq
\begin{aligned}
J_{\bigoplus_i R_i}(K;q) &= \sum_i J_{R_i}(K;q), \\
J_R(K^n;q) &= J_{R^{\otimes n}}(K;q).
\end{aligned}
\eeq
Here $K^n$ is the $n$-cabling of the knot~$K$, obtained by taking the path of~$K$ and tracing it with a ``cable'' of $n$ strands. These equations allow us to compute the $n$-colored Jones polynomial, given a way to compute the ordinary Jones polynomial and a little knowledge of representation theory. For instance, any knot~$K$ has $J_1(K;q) = 1$ and $J_2(K;q) = J (K;q)$, the ordinary Jones polynomial. Furthermore,
\beq
\begin{aligned}
\rep 2 \otimes \rep 2 = \rep 1 \oplus \rep 3
	&\implies J_3(K;q) = J(K^2;q) - 1, \\
\rep 2 \otimes \rep 2 \otimes \rep 2 = \rep 2 \oplus \rep 2 \oplus \rep 4
	&\implies J_4(K;q) = J(K^3;q) - 2 J(K;q),
\end{aligned}
\eeq
and so forth. We can switch to representations of lower dimension at the cost of considering more complicated links; however, the computability of the ordinary Jones polynomial means that this is still a good strategy for calculating colored Jones polynomials.

\begin{example}
Using the above formulae, it is easy to find $n$-colored Jones polynomial of the trefoil knot $K=3_1$
for the first few values of~$n$:
\beq
\begin{aligned}
J_1 & = 1, \\
J_2 & = q + q^3 - q^4, \\
J_3 & = q^2 + q^5 - q^7 + q^8 - q^9 - q^{10} + q^{11}, \\
& \vdots
\end{aligned}
\label{J-trefoil}
\eeq
where, for balance (and to keep the reader alert),
we used the conventions which differ from \eqref{trefJones} by the transformations \eqref{aqconf1} and \eqref{aqconf2}.
\end{example}

Much like the ordinary Jones polynomial is a particular specialization \eqref{JonesPq2} of the HOMFLY-PT polynomial,
its colored version $J_n(K; q)$ can be obtained by the same specialization from the so-called \vocab{colored HOMFLY-PT polynomial} $P_n (K;a,q)$,
\beq
J_n (K;q) = P_n (K;a=q^2,q).
\label{JonesPn}
\eeq
labeled by an integer $n$.
More generally, colored HOMFLY-PT polynomials $P^{\lambda} (K;a,q)$ are labeled by Young diagrams or 2d partitions $\lambda$.
In these lectures, we shall consider only Young diagrams that consist of a single row (or a single column)
and by Schur-Weyl duality correspond to totally symmetric (resp. totally anti-symmetric) representations.
Thus, what we call $P_n (K;a,q)$ is the HOMFLY-PT polynomial of $K$ colored by $\lambda = S^{n-1}$.

\begin{figure}
\includegraphics[width=5in]{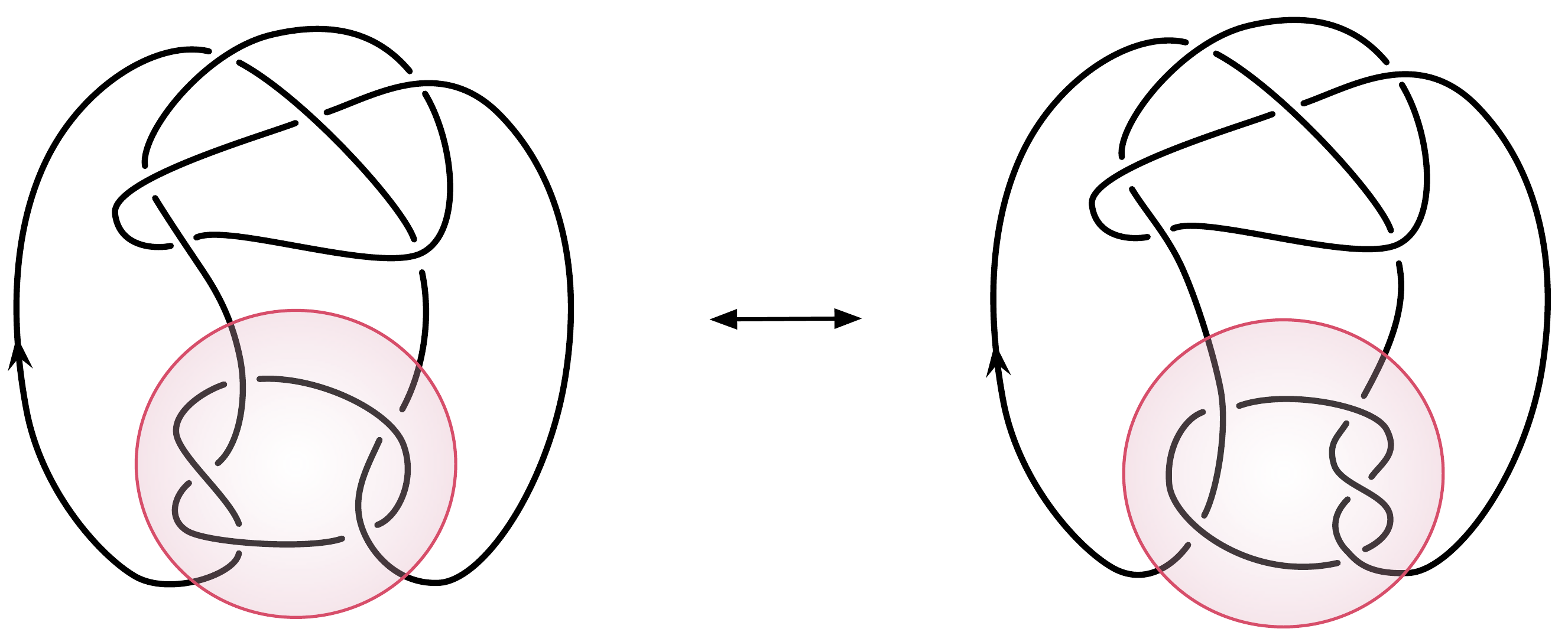}
\caption{Mutant knots. (Images from \cite{DGGindex}.)\label{fig:mutants}}
\end{figure}

Even though $P_n (K;a,q)$ provide us with an infinite sequence of two-variable polynomial knot invariants,
which can tell apart e.g. the two knots in \eqref{PP5110132},
they are still not powerful enough to distinguish simple pairs of knots and links called \vocab{mutants}.
The operation of \vocab{mutation} involves drawing a disc on a knot diagram such that two incoming and two outgoing strands pass its boundary, and then rotating the portion of the knot inside the disc by 180 degrees. The Kinoshita-Terasaka and Conway knots shown in Figure \ref{fig:mutants} are a famous pair of knots that are mutants of one another, but are nonetheless distinct; they can be distinguished by homological knot invariants, but not by any of the polynomial invariants we have discussed so far!

\begin{theorem}
The colored Jones polynomial, the colored HOMFLY-PT polynomial, and the Alexander polynomial
cannot distinguish mutants \cite{Morton}, while their categorifications can \cite{OSmutants,Wehrli,inprogress}.
\end{theorem}

\section{The classical $A$-polynomial}
\label{sec:Apol-1}

In this section, we take a step back from quantum group invariants to discuss another classical invariant of knots: the so-called \vocab{A-polynomial}. Our introduction will be rather brief, intended to familiarize the reader with the general idea behind this invariant and catalogue some of its properties, rather than attempt a complete construction. For more information, we refer to the pioneering paper of Cooper et.~al.~\cite{cooper}, in which the $A$-polynomial was first defined.

For a knot $K$, let $N(K)\subset S^3$ be an open tubular neighborhood of~$K$. Then the \vocab{knot complement} is defined to be
\beq M \defeq S^3 \setminus N(K). \label{knot-comp} \eeq
By construction, $M$ is a 3-manifold with torus boundary, and our goal here is to explain that to every such manifold
(not necessarily a knot complement) one can associate a planar algebraic curve
\beq
\mathscr{C} =  \{ (x,y) \in \C^2: A(x,y) = 0\}	\label{rep-variety},
\eeq
defined as follows. The classical invariant of $M$ is its fundamental group, $\pi_1(M)$,
which in the case of knot complements is called the \vocab{knot group}.
It contains a lot of useful information about $M$ and can distinguish knots much better than
any of the polynomial invariants we saw in section~\ref{sec:intro}.

\begin{example}
Consider the trefoil knot $K=3_1$. Its knot group is the simplest example of a braid group:
\beq \pi_1(M) = \langle a,b : aba=bab \rangle. \label{31knotgp} \eeq
\end{example}

\begin{figure}
\begin{tikzpicture}
	\draw[name path=middle,color=white,opacity=0] (2.5,0) -- (2.5,5);
	\draw[name path=top] (0,2) .. controls (0,4) and (5,4) .. (5,2);
	\draw[name path=bottom] (0,2) .. controls (0,0) and (5,0) .. (5,2);
	\draw[name path=htop] (1.5,1.8) .. controls (1.5,2.7) and (3.5,2.7) .. (3.5,1.8);
	\filldraw[color=white,fill=white,fill opacity=1] (1.5,2.2) .. controls (1.5,1.3) and (3.5,1.3) .. (3.5,2.2) -- (3.7,2.2)
	-- (3.5,1)
	-- (1.5,1) -- (1.3,2.2) -- (1.5,2.2);
	\draw[name path=hbottom] (1.5,2.2) .. controls (1.5,1.3) and (3.5,1.3) .. (3.5,2.2);
	\draw[name intersections={of=top and middle, by={a}}, name intersections={of=htop and middle, by={b}}]
		(a) .. controls (2.2,0|-a) and (2.2,0|-b) .. (b);
	\draw[name path=mright, name intersections={of=top and middle, by={a}}, name intersections={of=htop and middle, by={b}}, dashed]
		(a) .. controls (2.8,0|-a) and (2.8,0|-b) .. (b);
	\draw[name path=ltop, dashed] (0,2) .. controls (0,3.4) and (5,3.4) .. (5,2);
	\draw[name path=lbottom] (0,2) .. controls (0,0.5) and (5,0.5) .. (5,2);
	\fill[name intersections={of=ltop and mright, by={c}}, fill=gray] (c) circle (1.5pt);
	\draw (2.5,3.75) node {$m$};
	\draw[dotted,->,name intersections={of=lbottom and middle, by={d}}]
		(2.2,1.25) node[anchor=east] {$\ell$} .. controls (2.5,1.25) and (2.5,1.25) .. (d);
	\end{tikzpicture}
\caption{The torus $T^2 =\bdy N(K)$ for $K=\text{unknot}$, with cycles~$m$ and~$\ell$.\label{torus}}
\end{figure}
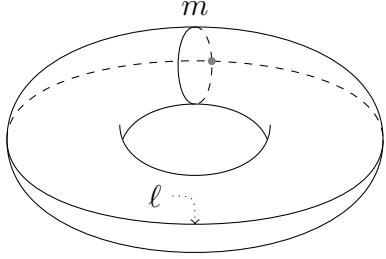

Although the knot group is a very good invariant, it is not easy to deal with due to its non-abelian nature.
To make life easier, while hopefully not giving up too much power,
one can imagine considering representations of the knot group rather than the group itself.
Thus, one can consider representations of $\pi_1(M)$ into a simple non-abelian group,
such as the group of $2 \times 2$ complex matrices,
\beq
\rho: \pi_1(M) \goesto \SL(2,\C).	\label{knotgp-reps} \eeq
Associated to this construction is a polynomial invariant $A(x,y)$, whose zero locus (\ref{rep-variety})
parameterizes in some sense the ``space'' of all such representations.
Indeed, as we noted earlier, $M$ is a 3-manifold with torus boundary,
\beq \bdy M = \bdy N(K) \isom T^2. \eeq
Therefore, the fundamental group of~$\bdy M$ is
\beq \pi_1(\bdy M) = \pi_1(T^2) = \Z\times\Z. \eeq
The generators of~$\pi_1(\bdy M)$ are the two basic cycles, which we will denote by~$m$ and$~\ell$ (standing for \vocab{meridian} and \vocab{longitude}, respectively---see Fig.~\ref{torus}). $m$ is the cycle that is contractible when considered as a loop in~$N(K)$,
and $\ell$ is the non-contractible cycle that follows the knot in~$N(K)$.
Of course, any representation $\pi_1(M) \goesto \SL(2,\C)$ restricts to a representation of~$\pi_1(T^2=\bdy M)$; this gives a natural map of representations of~$\pi_1(M)$ into the space of representations of~$\pi_1(\bdy M)$.

These cycles are represented in $\SL(2,\C)$ by $2 \times 2$ complex matrices $\rho(m)$ and~$\rho(\ell)$ with determinant~1.
Since the fundamental group of the torus is just $\Z\times\Z$, the matrices $\rho(m)$ and~$\rho(\ell)$ commute, and can therefore be simultaneously brought to Jordan normal form by some change of basis, i.e., conjugacy by an element of~$\SL(2,\C)$:
\beq
\rho(m) = \begin{pmatrix} x & \star \\ 0 & x^{-1} \end{pmatrix}, \quad \rho(\ell) = \begin{pmatrix} y & \star \\ 0 & y^{-1} \end{pmatrix}.
\label{matrices}
\eeq
Therefore, we have a map that assigns two complex numbers to each representation of the knot group:
\beq
\begin{aligned}
\Hom(\pi_1(M) , \SL(2,\C)) / \text{conj.} & ~~\goesto~~ \C^\star \times \C^\star , \\
\rho & ~~\mapsto~~ (x,y) ,
\end{aligned}
\label{rep-var-map}
\eeq
where~$x$ and $y$ are the eigenvalues of $\rho(m)$ and~$\rho(\ell)$, respectively. The image of this map is the \vocab{representation variety} $\mathscr{C}\subset \C^\star \times \C^\star$, whose defining polynomial is the $A$-polynomial of~$K$.
Note, this definition of the $A$-polynomial does not fix the overall numerical coefficient,
which is usually chosen in such a way that $A(x,y)$ has integer coefficients (we return to this property below).
For the same reason, the $A$-polynomial is only defined up to multiplication by arbitrary powers of $x$ and $y$.
Let us illustrate the idea of this construction with some specific examples.

\begin{example} \label{ex:A-unknot}
Let $K\subset S^3$ be the unknot. Then~$N(K)$ and~$M$ are both homeomorphic to the solid torus $S^1\times D^2$.
Notice that $m$ is contractible as a loop in~$N(K)$ and~$\ell$ is not, while the opposite is true in~$M$: $\ell$ is contractible and~$m$ is not.
Since~$\ell$ is contractible in~$M$, $\rho(\ell)$ must be the identity, and therefore we have $y=1$ for all $(x,y)\in\mathscr{C}$. There is no restriction on~$x$. Hence,
\beq
\mathscr{C}(\unknot) = \{ (x,y) \in \C^\star \times \C^\star : y=1 \},
\eeq
and the $A$-polynomial of the unknot is therefore
\beq
A(x,y) = y-1.
\eeq
\end{example}
\begin{example} \label{ex:A-trefoil}
Let $K\subset S^3$ be the trefoil knot $3_1$. Then, as mentioned in \eqref{31knotgp}, the knot group is given by
\beq
\pi_1(M) = \langle a,b : aba=bab \rangle,
\eeq
where the meridian and longitude cycles can be identified as follows:
\beq \begin{cases}
m=a, & \\
\ell = b a^2 b a^{-4}. & \end{cases} \label{trefoil-ml} \eeq
Let us see what information we can get about the $A$-polynomial just by considering abelian representations of~$\pi_1(M)$, i.e. representations such that $\rho (a)$ and $\rho (b)$ commute. For such representations, the defining relations reduce to $a^2 b = a b^2$ and therefore imply $a=b$. (Here, in a slight abuse of notation, we are simply writing $a$ to refer to $\rho(a)$ and so forth.) Eq.~\eqref{trefoil-ml} then implies that $\ell=1$ and $m=a$, so that $y=1$ and $x$ is unrestricted exactly as in Example~\ref{ex:A-unknot}. It follows that the $A$-polynomial contains $(y-1)$ as a factor.
\end{example}

This example illustrates a more general phenomenon. Whenever~$M$ is a knot complement in~$S^3$, it is true that the abelianization
\beq
\pi_1(M)^\text{ab} = H_1(M) \isom \Z.
\eeq
Therefore, the $A$-polynomial always contains $y-1$ as a factor,
\beq
A(x,y) = (y-1)(\cdots),
\eeq
where the first piece carries information about abelian representations, and any additional factors that occur arise from the non-abelian representations. In the particular case $K=3_1$, a similar analysis of non-abelian representations of (\ref{31knotgp}) into $\SL(2,\C)$ yields
\beq
A(x,y) = (y-1) (y+x^6).	\label{eq:A-trefoil}
\eeq

To summarize, the algebraic curve $\mathscr{C}$ is (the closure of) the image of the representation variety of~$M$ in the representation variety $\C^\star \times \C^\star$ of its boundary torus~$\bdy M$. This image is always an affine algebraic variety of complex dimension 1, whose defining equation is precisely the $A$-polynomial~\cite{cooper}.

This construction defines the $A$-polynomial as an invariant associated to any knot.
However, extension to links requires extra care, since in that case $\bdy N(L)\not\isom T^2$. Rather, the boundary of the link complement consists of several components, each of which is separately homeomorphic to a torus. Therefore, there will be more than two fundamental cycles to consider, and the analogous construction will generally produce a higher-dimensional character variety rather than a plane algebraic curve. One important consequence of this is that the $A$-polynomial cannot be computed by any known set of skein relations; as was made clear in Exercise~\ref{hw:HOMFLY}, computations with skein relations require one to consider general links rather than just knots.

To conclude this brief introduction to the $A$-polynomial, we will list without proof several of its interesting properties:
\begin{itemize}
\item For any hyperbolic knot $K$,
\beq
A_K (x,y) \neq y-1.
\eeq
That is, the $A$-polynomial carries nontrivial information about non-abelian representations of the knot group.
\item
Whenever $K$ is a knot in a homology sphere, $A_K(x,y)$ contains only even powers of the variable $x$.
Since in these lectures we shall only consider examples of this kind, we simplify expressions a bit by replacing $x^2$ with $x$.
For instance, in these conventions the $A$-polynomial (\ref{eq:A-trefoil}) of the trefoil knot looks like
\beq
A(x,y) = (y-1) (y+x^3).	\label{eq:A-trefoil1}
\eeq
\item The $A$-polynomial is \vocab{reciprocal}: that is,
\beq
A(x,y) \sim A(x^{-1},y^{-1}),
\eeq
where the equivalence is up to multiplication by powers of $x$ and~$y$. Such multiplications are irrelevant, because they don't change the zero locus of the $A$-polynomial in $\C^\star \times \C^\star$. This property can be also expressed by saying that the curve~$\mathscr{C}$ lies in $(\C^\star \times \C^\star) / \Z_2$, where $\Z_2$ acts by $(x,y)\mapsto (x^{-1},y^{-1})$ and can be interpreted as the Weyl group of~$\SL(2,\C)$.
\item
$A(x,y)$ is invariant under orientation reversal of the knot, but \textsl{not} under reversal of orientation in the ambient space. Therefore, it can distinguish mirror knots (knots related by the parity operation), such as the left- and right-handed versions of the trefoil. To be precise, if~$K'$ is the mirror of~$K$, then
\beq
A_K (x,y) = 0 \iff A_{K'} (x^{-1},y) = 0.
\eeq
\item
After multiplication by a constant, the $A$-polynomial can always be taken to have integer coefficients.
It is then natural to ask: are these integers counting something, and if so, what?
The integrality of the coefficients of $A(x,y)$ is a first hint of the deep connections with number theory.
For instance, the following two properties, based on the \vocab{Newton polygon} of $A(x,y)$, illustrate this connection further.
\item
The $A$-polynomial is \vocab{tempered}: that is, the faces of the Newton polygon of $A(x,y)$ define cyclotomic polynomials in one variable.
Examine, for example, the $A$-polynomial of the figure-8 knot:
\beq
A(x,y) = (y-1) \left( y^2 - (x^{-2} -x^{-1} -2 - x + x^2) y + y^2 \right) .
\label{eq:A-41}
\eeq
\item
Furthermore, the slopes of the sides of the Newton polygon of $A(x,y)$ are boundary slopes of incompressible
surfaces\footnote{A proper embedding of a connected orientable surface $F \goesto M$ is called incompressible if
the induced map $\pi_1 (F) \goesto \pi_1 (M)$ is injective. Its boundary slope is defined as follows.
An incompressible surface $(F, \bdy F)$ gives rise to a collection of parallel simple closed loops in $\bdy M$.
Choose one such loop and write its homology class as $\ell^p m^q$. Then, the boundary slope of $(F, \bdy F)$
is defined as a rational number $p/q$.} in $M$.
\end{itemize}

While all of the above properties are interesting, and deserve to be explored much more fully,
our next goal is to review the connection to physics \cite{gukov},
which explains known facts about the $A$-polynomial and leads to many new ones:
\begin{itemize}

\item
The $A$-polynomial curve (\ref{rep-variety}), though constructed as an algebraic curve, is most properly viewed as an object of symplectic geometry: specifically, a holomorphic Lagrangian submanifold.

\item
Its quantization with the symplectic form
\beq
\omega=\frac{dy}{y} \wedge \frac{dx}{x}
\label{sympformxy}
\eeq
leads to interesting wavefunctions.

\item
The curve~$\mathscr{C}$ has all the necessary attributes to be an analogue of the Seiberg-Witten curve for knots and $3$-manifolds \cite{DGH,FGSS}.

\end{itemize}

As an appetizer and a simple example of what the physical interpretation of the $A$-polynomial has to offer,
here we describe a curious property of the $A$-polynomial curve (\ref{rep-variety}) that follows from this physical interpretation.
For any closed cycle in the algebraic curve~$\mathscr{C}$, the integral of the Liouville one-form
associated to the symplectic form \eqref{sympformxy} should be quantized \cite{gukov}.
Schematically,\footnote{To be more precise, all periods of the ``real'' and ``imaginary'' part
of the Liouville one-form $\theta$ must obey
$$
\begin{aligned}
& \oint_{\Gamma} \Big( \log |x| d ({\rm arg} \, y) - \log |y| d ({\rm arg} \, x) \Big) \; = \; 0 \,, \\
\frac{1}{4 \pi^2} & \oint_{\Gamma} \Big( \log |x| d \log |y| + ({\rm arg} \, y) d ({\rm arg} \, x) \Big) \; \in \; \mathbb{Q} \,.
\end{aligned}
$$
}
\beq
\oint_\Gamma \log y \frac{dx}{x} \in 2\pi^2 \cdot \Q.
\label{integrality}
\eeq
This condition has an elegant interpretation in terms of algebraic $K$-theory and the Bloch group of~$\bar{\Q}$.
Moreover, it was conjectured in \cite{abmodel} that every curve of the form (\ref{rep-variety})
--- not necessarily describing the moduli of flat connections ---
is quantizable if and only if $\{ x,y \} \in K_2(\C(\mathscr{C}))$ is a torsion class.
This generalization will be useful to us later, when we consider a refinement of the $A$-polynomial
that has to do with categorification and homological knot invariants.

To see how stringent the condition \eqref{integrality} is, let us compare, for instance,
the $A$-polynomial of the figure-eight knot \eqref{eq:A-41}:
\beq
A(x,y) = 1- ( x^{-4} - x^{-2} - 2 - x^2 + x^4) y + y^2
\eeq
and a similar polynomial
\beq
B(x,y) = 1- ( x^{-6} - x^{-2} - 2 - x^2 + x^6) y + y^2 \,.
\eeq
(Here the irreducible factor $(y-1)$, corresponding to abelian representations, has been suppressed in both cases.)
The second polynomial has all of the required symmetries of the $A$-polynomial,
and is obtained from the $A$-polynomial of the figure-eight knot by a hardly noticeable modification.
But $B(x,y)$ cannot occur as the $A$-polynomial of any knot since it violates the condition \eqref{integrality}.

\section{Quantization}	\label{sec:quantization}	\index{quantization}

Our next goal is to explain, following \cite{gukov}, how physical interpretation of the $A$-polynomial
in Chern-Simons theory can be used to provide a bridge between \textsl{quantum group invariants} of knots and \textsl{algebraic curves}
that we discussed in sections \ref{sec:intro} and \ref{sec:Apol-1}, respectively.
In particular, we shall see how quantization of Chern-Simons theory
naturally leads to a \vocab{quantization} of the classical curve (\ref{rep-variety}),
\beq
A(x,y) \quad \leadsto \quad \op A(\op x, \op y; q) \,,
\label{Aquantproc}
\eeq
i.e. a $q$-difference operator $\op A(\op x, \op y; q)$ with many interesting properties.
While this will require a crash course on basic tools of Quantum Mechanics, the payoff will be enormous
and will lead to many generalizations and ramifications of the intriguing relations
between quantum group invariants of knots, on the one hand, and algebraic curves, on the other.
Thus, one such generalization will be the subject of section \ref{sec:categorification},
where we will discuss categorification and formulate a similar bridge between algebraic curves
and \textsl{knot homologies}, finally explaining the title of these lecture notes.\\

We begin our discussion of the quantization problem with a lightning review of some mathematical aspects of classical mechanics.
Part of our exposition here follows the earlier lecture notes~\cite{dimofte}
that we recommend as a complementary introduction to the subject.
When it comes to Chern-Simons theory, besides the seminal paper \cite{WittenCS},
mathematically oriented readers may also want to consult excellent books~\cite{atiyah,kohno}.

As we discussed briefly in the introduction,
the description of a system in classical mechanics is most naturally formulated in the language of symplectic geometry.
In the classical world, the state of a system at a particular instant in time is completely specified by giving $2N$ pieces of data: the values of the coordinates~$x_i$ and their conjugate momenta~$p_i$, where $1\leq i \leq N$. The $2N$-dimensional space parameterized by the~$x_i$ and~$p_i$ is the \vocab{phase space}~$\moduli$ of the system. (For many typical systems, the space of possible configurations of the system is some manifold~$X$, on which the~$x_i$ are coordinates, and the phase space is the cotangent bundle $\moduli=\ctb\, X$.)
Notice that, independent of the number~$N$ of generalized coordinates needed to specify the configuration of a system, the associated phase space is always of even dimension. In fact, phase space is always naturally equipped with the structure of a symplectic manifold, with a canonical symplectic form given by
\beq
\omega = dp \wedge dx.	\label{symplectic-form}
\eeq
(When the phase space is a cotangent bundle, \eqref{symplectic-form} is just the canonical symplectic structure on any cotangent bundle, expressed in coordinates.) Recall that a symplectic form on a manifold is a closed, nondegenerate two-form, and that nondegeneracy immediately implies that any symplectic manifold must be of even dimension.

Since $\omega$ is closed, locally it admits a primitive form, the so-called \vocab{Liouville one-form}
\beq
\theta = p\, dx.
\eeq
It should be apparent that $\omega = d \theta$, so that $\theta$ is indeed a primitive.

\begin{figure}
\includegraphics[height=1.5in]{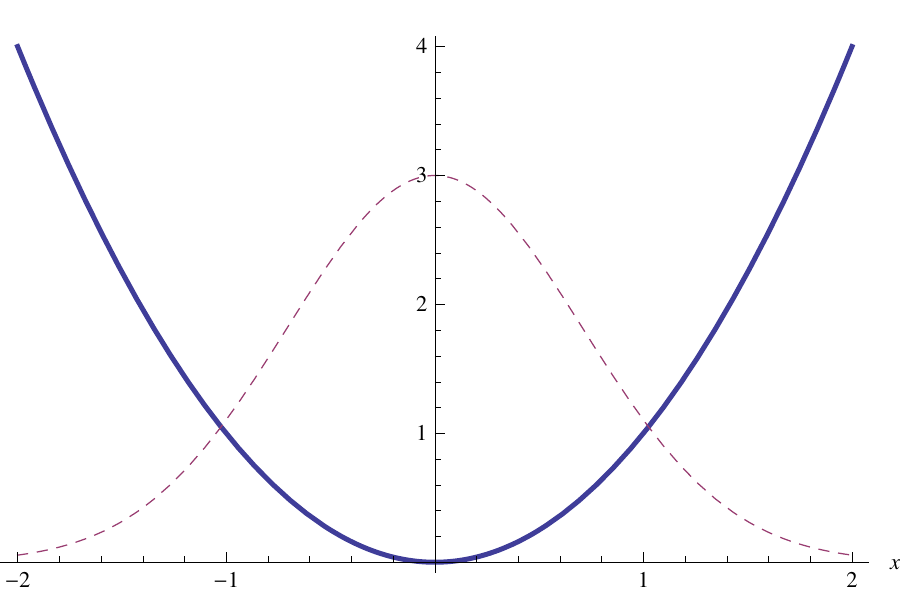}
\includegraphics[height=1.5in]{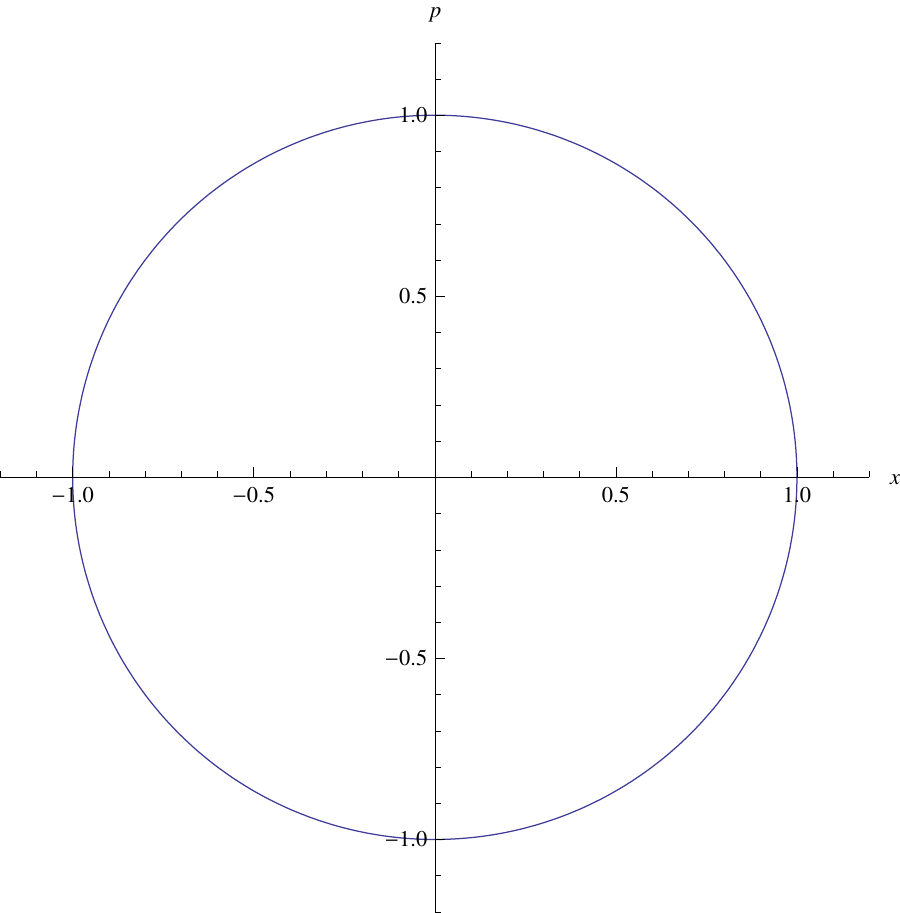}
\caption{On the left, the potential and lowest-energy wavefunction for the simple harmonic oscillator. On the right, the phase space of this system, with a typical classical trajectory.\label{fig:SHO}}
\end{figure}

Let us now explore these ideas more concretely in the context of a simple example. As a model system, consider the one-dimensional simple harmonic oscillator. The configuration space of this system is just $\R$ (with coordinate~$x$), and the Hamiltonian is given by
\beq
H = \frac{1}{2} p^2 + \frac{1}{2} x^2.
\eeq
Since $dH/dt = 0$, the energy is a conserved quantity, and since $N=1$, this one conserved quantity serves
to completely specify the classical trajectories of the system. They are curves in phase space of the form
\beq
\mathscr{C} : \frac{1}{2} (x^2 + p^2) - \E = 0,	\label{eq:trajectories}
\eeq
for $\E \in \R_+$; these are concentric circles about the origin, with radius determined by the energy. Figure~\ref{fig:SHO} shows the potential of this system, together with a typical trajectory in the phase space. The dashed line represents the lowest-energy wavefunction of the system, to which we will come in a moment.

Now, recall that a \vocab{Lagrangian submanifold} $\mathscr{C} \subset (\moduli,\omega)$ is a submanifold such that $\omega|_{\mathscr{C}}=0$, having the maximal possible dimension, i.e., $\dim \mathscr{C} = \frac{1}{2} \dim \moduli$. (If~$\mathscr{C}$ has dimension larger than half the dimension of~$\moduli$, the symplectic form cannot be identically zero when restricted to~$\mathscr{C}$, since it is nondegenerate on~$\moduli$.)
It should be clear that, in the above example, the classical trajectories~\eqref{eq:trajectories} are {Lagrangian submanifolds} of the phase space.

Moreover, since in this example the degree of the symplectic form~$\omega$ is equal to the dimension of the phase space, $\omega$ is a volume form --- in fact, the standard volume form on~$\R^2$. We can therefore compute the area encompassed by a trajectory of energy~$\E$ by integrating~$\omega$ over the region $x^2 + p^2 < 2\E$, obtaining
\beq
2\pi \E = \int_D dp\wedge dx,
\eeq
where $D$ is the disc enclosed by the trajectory~$\mathscr{C}$. Therefore, classically, the energy of a trajectory is proportional to the area in phase space it encompasses.

How do these considerations relate to quantization of the system? It is well known that the energy levels of the simple harmonic oscillator are given by
\beq
\E = \frac{1}{2\pi} \int_D dp\wedge dx
= \hbar \left(n + \frac{1}{2}\right)
\eeq
when the system is quantized. Thus, we expect that, in quantizing a system, the number of quantum states contained in some region of phase space will be directly proportional to its area. Moreover, we interpret $\hbar$, which has the same units as area in phase space, as the amount of classical phase space per quantum state. Schematically,
\beq
\# \text{ states} \sim \text{area}/\hbar.
\eeq
This relation has a long history in quantum physics; it is none other than the Bohr-Sommerfeld quantization condition.

Moreover, since $\omega$ admits a primitive, we can use the Stokes theorem to write
\beq
\E = \frac{1}{2\pi} \int_D \omega = \frac{1}{2\pi} \oint_\mathscr{C} \theta,	\label{eq:int1}
\eeq
since $\mathscr{C} = \bdy D$ and $d\theta = \omega$.

We have discussed counting quantum states; what about actually constructing them? In quantum mechanics, we expect the state to be a vector in a Hilbert space, which can be represented as a square-integrable wavefunction $Z(x)$. It turns out that, in the limit where $\hbar$ is small, the wavefunction can be constructed to lowest order in a manner that bears a striking resemblance to~\eqref{eq:int1}:
\beq
\begin{aligned}
Z(x)  \xrightarrow[\hbar\goesto 0]{} & \exp\left[ \frac{i}{\hbar} \int_0^x \theta + \cdots \right] \\
= & \exp \left[ \frac{i}{\hbar} \int_0^x \sqrt{2\E - x^2}\, dx + \cdots \right] \\
\end{aligned}
\eeq
Evaluating the wavefunction in this manner for the lowest-energy state of our system ($\E = \hbar/2$) yields
\beq
Z(x) \approx \exp \left[-\frac{1}{2\hbar} x^2 + \cdots \right].
\eeq
Indeed, $exp(-\frac{x^2}{2\hbar})$ is the exact expression for the $n=0$ wavefunction.

We are slowly making progress towards understanding the quantization of our model system. The next step is to understand the transition between the classical and quantum notions of an \vocab{observable}. In the classical world, the observables~$x$ and~$p$ are coordinates in phase space --- in other words, functions \textsl{on} the phase space:
\beq x: \moduli \goesto \R, \
(x,p) \mapsto x, \eeq
and so forth. General observables are functions of~$x$ and~$p$, i.e., general elements of~$\Cinf(\moduli,\R)$.

In the quantum world, as is well known, $x$ and~$p$ should be replaced by operators $\op x$ and~$\op p$, obeying the canonical commutation relation
\beq [ \op p, \op x ] = -i\hbar. 	\label{commrel}	\eeq
These operators now live in some noncommutative algebra, which is  equipped with an action on the Hilbert space of states. In the position representation, for instance,
\beq \op x f(x) = x f(x), \quad \op p f(x) = - i \hbar \frac{d}{dx} f(x), \eeq
where $f\in L^2(\R)$. The constraint equation~\eqref{eq:trajectories} that defines a classical trajectory is then replaced by the operator equation
\beq
\left[ \frac{1}{2} (\op x^2 + \op p^2 ) - \E \right] Z(x) = 0, \label{schrodinger}
\eeq
which is just the familiar Schr\"odinger eigenvalue equation $\op H Z = \E Z$. Now, unlike in the classical case, the solutions of~\eqref{schrodinger} in the position representation will only be square-integrable (and therefore physically acceptable) for certain values of~$\E$. These are precisely the familiar eigenvalues or allowed energy levels
\beq
\E = \hbar \left( n+ \frac{1}{2} \right) .
\eeq
Taking $n=0$, for example, the exact  solution is $Z(x)=\exp(-x^2/2\hbar)$, just as we claimed above, as the reader may easily verify.

All of this discussion should be taken as illustrating our above claim that quantum mechanics
should properly be understood as a ``modern symplectic geometry,''
in which classical constraints are promoted to operator relations.
We have constructed the following correspondence or dictionary between the elements of the classical and quantum descriptions of a system:

\medskip
\begin{center} {\small
\begin{tabular}{|l|c|c|} \hline
	& {\sl Classical} & {\sl Quantum} \\ \hline
state space		& symplectic manifold $(\moduli,\omega)$	& Hilbert space $\Hil$ \\	\hline
states		& Lagrangian submanifolds 			& vectors (wave functions) \\
~~	&	$\mathscr{C} \subset \moduli$					&  $Z\in\Hil$ \\	\hline
observables		& algebra of functions 				& algebra of operators \\
			&$f \in\Cinf(M)$			& $\op f$, acting on~$\Hil$ \\	\hline
constraints & $f_i =0 $ & $\op f_i Z = 0$ \\ \hline
\end{tabular}}	\end{center}
\medskip

\noindent
We now have a benchmark for what a successful quantization should accomplish: for a given classical system, it should construct the quantum counterpart for each element in the classical description, as summarized above. Of course, we would also like the correspondence principle to hold: that is, the quantum description should dovetail nicely with the classical one in some way when we take $\hbar\goesto 0$.

The correspondence between the classical and quantum descriptions is not quite as cut-and-dried as we have made it appear, and there are a few points that deserve further mention. Firstly, it should be apparent from our discussion of the harmonic oscillator that not every Lagrangian submanifold will have a quantum state associated to it; in particular, only a particular subset of these (obeying the Bohr-Sommerfeld quantization condition, or equivalently, corresponding to eigenvalues of the operator~$\op H$) will allow us to construct a square-integrable wavefunction~$Z(x)$.
There can be further constraints on \vocab{quantizable} Lagrangian submanifolds \cite{branequant}.

Secondly, let us briefly clarify why quantum state vectors correspond to Lagrangian submanifolds of the classical phase space and not to classical 1-dimensional trajectories, as one might naively think. (In our example of the harmonic oscillator we have $N=1$ and, as a result, both Lagrangian submanifolds and classical trajectories are one-dimensional.) The basic reason why Lagrangian submanifolds, rather than dimension-1 trajectories, are the correct objects to consider in attempting a quantization is the following.
In quantum mechanics, we specify a state by giving the results of measurements of observables performed on that state. For this kind of information to be meaningful, the state must be a simultaneous eigenstate of all observables whose values we specify, which is only possible if all such observables mutually commute. As such, to describe the state space in quantum mechanics, we choose a ``complete set of commuting observables'' that gives a decomposition of~$\Hil$ into one-dimensional eigenspaces of these operators. For time-independent Hamiltonians, one of these operators will always be~$\op H$.

However, to the leading order in $\hbar$ the commutator of two quantum observables must be proportional to the Poisson bracket of the corresponding classical observables. Therefore, if $\op H, \op f_i$ form a complete set of commuting quantum-mechanical observables,
we must have
\beq
\{ H, f_i \} _\text{P.B.} =0,
\eeq
where $\{ \cdot\, , \cdot \}_\text{P.B.}$ is the Poisson bracket. But we know that the classical time-evolution of the quantity~$f_i$ is determined by the equation
\beq
\frac{df_i}{dt} + \{ H, f_i \}_\text{P.B.} = 0.
\eeq
As such, the quantum-mechanical observables used in specifying the state must correspond to classically conserved quantities: \vocab{constants of the motion}. And it is well-known that the maximal possible  number of classically conserved quantities is $N = \frac{1}{2} \dim \moduli$, corresponding to a completely integrable system; this follows from the nondegeneracy of the symplectic form on the classical phase space. For $N>1$, then, specifying all of the constants of the motion does not completely pin down the classical trajectory; it specifies an $N$-dimensional submanifold $\mathscr{C} \subset \moduli$. However, it does give all the information it is possible for one to have about the quantum state. This is why Lagrangian submanifolds are the classical objects to which one attempts to associate quantum states.

We should also remark that it is still generically true that wavefunctions will be given to lowest order by
\beq
Z(x) = \exp \left[ \frac{i}{\hbar} \int_{x_0}^x \theta + \cdots \right].
\eeq
This form fits all of the local requirements for~$Z(x)$, although it may or may not produce a globally square-integrable wavefunction.

Finally, the quantum-mechanical algebra of operators is a non-commutative deformation or \vocab{$q$-deformation} of the algebra of functions $\Cinf(\moduli)$, where the deformation is parameterized by
\beq
q \defeq e^{\hbar}.
\eeq
In the classical limit, $q\goesto 1$.

How does the idea of quantization bear any relation to the ostensible subject of this lecture series, topological quantum field theories? To illustrate the connection, we will consider a specific example of a TQFT: the Chern-Simons gauge theory.
\index{Chern-Simons theory}

As in any gauge theory, the starting point of this theory is the choice of a gauge group~$G$
and the action functional, which in the present case is the Chern-Simons functional:
\beq
\label{CS-action}
\frac{1}{\hbar} \int_M \Tr ( A \wedge dA + \frac{2}{3} A \wedge A \wedge A) .
\eeq
Here $M$ is a 3-manifold, and the gauge field~$A$ is a connection on a principal $G$-bundle $E\goesto M$. The action functional~\eqref{CS-action} can be interpreted roughly as a Morse function on the space of gauge fields. We search for critical points of this functional by solving the equation of motion, which is the PDE
\beq
\label{CS-EOM}
dA + A\wedge A = 0.
\eeq
This equation says that~$A$ is a flat connection. How is this gauge theory formulation related to the picture of a TQFT as a functor, in the axiomatic language of Atiyah and Segal \cite{atiyah}?

The answer to this question is summarized in the below table, and illustrates the way in which quantization plays a role. The action functional~\eqref{CS-action} defines a classical gauge theory; the classical phase space of this theory is the \vocab{moduli space of flat connections} $\moduli = \moduli_\text{flat}(G, \Surf)$, where $\Surf=\bdy M$.

\begin{figure}
\begin{tikzpicture}
	\draw (5,5) .. controls (2.5,5) and (4,3) .. (4,2.5);
	\draw (5,5) .. controls (7.5,5) and (6,3) .. (6,2.5);
	\draw (5,0) .. controls (2.5,0) and (4,2) .. (4,2.5);
	\draw (5,0) .. controls (7.5,0) and (6,2) .. (6,2.5);
	\draw (4.85,4.2) .. controls (5.45,4.2) and (5.45,3.2) .. (4.85,3.2);
	\filldraw[fill=white, color=white]
		(5.15,4.2) .. controls (4.55,4.2) and (4.55,3.2) .. (5.15,3.2)
		-- (5.15,3) -- (3.7,3.7) -- (5.15,4.4) -- (5.15,4.2);
	\draw (5.15,4.2) .. controls (4.55,4.2) and (4.55,3.2) .. (5.15,3.2);
	\draw (4.85,0.8) .. controls (5.45,0.8) and (5.45,1.8) .. (4.85,1.8);
	\filldraw[fill=white, color=white]
		(5.15,0.8) .. controls (4.55,0.8) and (4.55,1.8) .. (5.15,1.8)
		-- (5.15,2) -- (3.7,1.3) -- (5.15,0.6) -- (5.15,0.8);
	\draw (5.15,0.8) .. controls (4.55,0.8) and (4.55,1.8) .. (5.15,1.8);
	\draw (5,2.5) node{$\Surf$};
	\draw (5,5) .. controls (3.5,5) and (3.5,4.5) .. (3,4.5);
	\draw (3,4.5) .. controls (2,4.5) and (2,5) .. (1,5);
	\draw (5,0) .. controls (3.5,0) and (3.5,0.5) .. (3,0.5);
	\draw (3,0.5) .. controls (2,0.5) and (2,0) .. (1,0);
	\draw (1,0) .. controls (-1,0) and (-1,5) .. (1,5);
	\draw (1.8,2.5) node{$M$};
\end{tikzpicture}
\caption{The setup for Chern-Simons theory: an oriented 3-manifold $M$ with boundary a 2-manifold~$\Surf$.\label{3manifold}}
\end{figure}
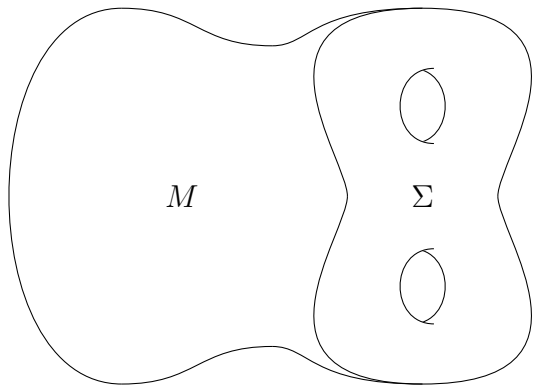

Now, let $\moduli_\text{flat}(G,M)$ be the moduli space of flat connections on~$M$. There is a natural mapping
\beq
\moduli_\text{flat}(G,M) \inj \moduli_\text{flat}(G,\Surf)
\eeq
induced by restriction to~$\Surf = \bdy M$. The image of this map is the subspace of~$\moduli$ consisting of flat connections on~$\Surf$ that can be extended to~$M$; this is a Lagrangian submanifold $\mathscr{C}\subset\moduli$.

We are now equipped with precisely the classical data referred to in our earlier discussion of the quantization problem. If we now quantize the classical Chern-Simons theory, the classical phase space~$\moduli$ and the Lagrangian submanifold~$\mathscr{C}\subset \moduli$ will be respectively replaced with a Hilbert space and a state vector in that Hilbert space. But these are precisely the objects that we expect a TQFT functor to associate to~$\Surf$ and~$M$!

To sum up, our situation is as follows: \medskip

\begin{tabular}{|l|l|l|} \hline
{\sl Geometry} & {\sl Classical CS} & {\sl Quantum CS} \\ \hline
2-manifold & symplectic manifold & vector space \\
   $\Surf$ & $\moduli = \moduli_\text{flat}(G,\Surf)$ & $\Hil_\Surf$ \\ \hline
3-manifold	& Lagrangian submanifold: & vector \\
$M$	($\bdy M = \Surf$)	&  connections extendible to~$M$	& $Z(M) \in \Hil_\Surf$ \\ \hline
\end{tabular} \medskip

\noindent
To move from the first column to the second, we define the classical Chern-Simons theory. Moving from the second column to the third consists of a quantization of this theory. The usual picture of a TQFT as a functor is the composition of these two: it moves directly from the first to the third column, ignoring the second.


Let us discuss the phase space of classical Chern-Simons theory a little further. It is known that all flat connections on Riemann surfaces are described by their holonomies; that is, the moduli space consists of maps
\beq
\moduli = \Hom ( \pi_1(\Surf) \goesto G) / \text{conjugation}.
\eeq
As emphasized in the work of Atiyah and Bott~\cite{AB},
this space comes equipped with a natural symplectic form,
\beq
\omega = \frac{1}{4\pi^2} \int_\Surf \Tr \delta A \wedge \delta A,
\label{moduli-symplectic}
\eeq
where $\delta$ denotes the exterior derivative on $\moduli$, so that $\delta A$ is a 1-form on $\Surf$ as well as on $\moduli$.
The Lagrangian submanifold we are considering is then given by
\beq
\mathscr{C} = \Hom(\pi_1(M) \goesto G) / \text{conjugation},
\eeq
and the inclusion map is induced by the natural map $\pi_1(\Surf)\goesto \pi_1(M)$.
This Lagrangian submanifold can be defined by classical constraint equations of the form
\beq
A_i = 0.
\eeq
Quantization will then replace these with quantum constraints; that is, operator relations
\beq
\op A_i Z = 0 \label{op-zero}
\eeq
much like the classical constraint \eqref{eq:trajectories} was replaced by the operator equation \eqref{schrodinger} in our previous example.

\begin{exercise}
Verify that $\mathscr C$ is indeed Lagrangian with respect to the symplectic form~\eqref{moduli-symplectic}.
That is, show that
\beq
\omega|_{\mathscr{C}\subset\moduli} = 0.
\eeq
\end{exercise}

\begin{exercise}
Let~$g$ be the genus of~$\Surf$. Show that, for $g>1$,
\beq
\dim \moduli = (2g-2) \dim G.
\eeq
\begin{proof}[Solution]
Consider the case where~$G$ is a simple group.
The fundamental group $\pi_1(\Surf)$ is generated by $2g$ elements $A_i$ and~$B_i$, $1\leq i \leq g$, subject to the one relation
\beq
A_1 B_1 A_1^{-1} B_1^{-1} 
\cdots  A_g B_g A_g^{-1} B_g^{-1} = 1. \label{fundgp-relation}
\eeq
After applying an element of $\Hom(\pi_1(\Surf) \goesto  G)$, the generators are mapped to group-valued matrices, and so we need $2g\cdot \dim G$ parameters to specify them all independently. However, there are constraints: the matrices must obey~\eqref{fundgp-relation}, one matrix equation which eliminates $\dim G$ degrees of freedom. Taking the quotient by conjugacy eliminates another $\dim G$ degrees of freedom, leaving
\beq
\dim\moduli = (2g -2 ) \dim G,
\eeq
as we expected.
\end{proof}
\end{exercise}

Let us now specialize this general discussion and consider the theory with gauge group $G = \SL(2,\C)$ on a 3-manifold that is a knot complement, $M=S^3 \setminus N(K)$. Then, of course, $\bdy M = \Surf \isom T^2$. It follows immediately that $\pi_1(\Surf) = \Z\times\Z$,  so that
\beq
\begin{aligned}
\moduli &= \Hom(\Z\times\Z \goesto \SL(2,\C))/\text{conjugacy} \\
&= (\C^\star \times \C^\star) / \Z_2.
\end{aligned}
\eeq
This is exactly the space we considered in section \ref{sec:Apol-1} in our discussion of the $A$-polynomial:
it is the representation variety of the boundary torus of~$M$! Moreover, the Lagrangian submanifold is in this case given by
\beq
\begin{aligned}
\mathscr{C} &= \Hom(\pi_1(M)\goesto\SL(2,\C))/\text{conjugacy} \\
&= \{ (x,y) \in (\C^\star \times \C^\star) / \Z_2: A(x,y) = 0\},
\end{aligned}
\label{ACurve}
\eeq
where $A(x,y)$ is a familiar polynomial in $x$ and~$y$, interpreted now as a classical observable giving the classical constraint relation that defines the submanifold $\mathscr{C}\subset\moduli$.

The appearance of the $A$-polynomial in this context clarifies two mysterious statements that were made in the previous section. Firstly, it makes apparent in what sense the zero locus of the $A$-polynomial is a natural object in symplectic geometry. Secondly, we can now make sense of the statement that one can ``quantize'' the $A$-polynomial. Having interpreted it as a classical constraint equation defining a Lagrangian submanifold of a classical phase space, it becomes obvious that quantization replaces the $A$-polynomial by an operator in a quantum constraint equation of the form~\eqref{op-zero}.

What happens when we try to quantize the $A$-polynomial? The natural symplectic form~\eqref{moduli-symplectic} on the classical phase space takes the simple form \cite{gukov}:
\beq
\omega = \frac{dy}{y}  \wedge \frac{dx}{x}  = d\ln y \wedge d\ln x.
\label{xysympform}
\eeq
The canonical commutation relation is therefore
\beq
\left[ \widehat{\ln y}, \widehat{\ln x} \right] = \hbar,
\eeq
which can be rewritten in the form
\beq
\op y \op x =  q \op x \op y .
\eeq
with $q=e^\hbar$. Given this relation, what form do the operators $\op x$ and~$\op y$ take in the position representation? Of course, we must have $\op x f(x) = x f(x)$. Then the commutation relation becomes
\beq
q \op x (\op y f(x)) = \op y (\op x f(x)),
\eeq
and implies that $\op y$ should act as a \vocab{shift operator} $\op y f(x) = f(qx)$.
The reason for this name is the following.
Notice, that the symplectic form \eqref{xysympform} has the canonical form in logarithms of $x$ and $y$,
rather than $x$ and $y$ themselves. Therefore, it is natural to introduce the logarithmic variable $n$ by the relation $x=q^n$.
Then, in terms of $n$ the action of the operators $\op x$ and $\op y$ looks like
\beq
\op x f(n) = q^n f(n), \quad \op y f(n) = f(n+1).
\label{xnyshift}
\eeq
The quantization of the polynomial $A(x,y) = \sum_k a_k(x) y^k$ will then be an operator of the form
\beq
\op A(\op x, \op y; q) = \sum_k  a_k(\op x;q) \op y^k. \label{Aopquant}
\eeq

In general, quantization is a rather delicate and mysterious procedure \cite{woodhouse} (see \cite{branequant} for a recent discussion).
However, for algebraic curves defined by classical constraint equations of the form $A(x,y)=0$,
recent progress in mathematical physics  \cite{Mironov,Marino:2006hs,eyn-or,BKMP,DijkgraafFuji-2} has led to a systematic way
of constructing the coefficients $a_k(\op x;q)$ of the quantum operator \eqref{Aopquant} entirely from the data
of the classical $A$-polynomial \cite{abmodel} (see also \cite{BorotE}):
\beq
A(x,y) \quad \leadsto \quad \op A(\op x, \op y; q). 
\label{quantAproc}
\eeq
In addition, in some cases the curve $A(x,y) = 0$ comes from extra data that may be very helpful in constructing
its quantum analog. For instance, the construction \cite{NZ} of the ordinary $A$-polynomial
based on the triangulation data of a 3-manifold $M$ admits a beautiful non-commutative lift \cite{Tudor}.
However, since in what follows we need to apply the procedure \eqref{quantAproc} to arbitrary curves for which
the extra data is not always available, we shall mainly focus on the so-called \vocab{topological recursion}
approach that involves complex analysis and noncommutative algebra on $\mathscr{C}$.

In complex analysis, one of the basic ingredients associated to the curve $\mathscr{C} : A(x,y) = 0$ is the so-called \vocab{Bergman kernel}.
It becomes the first brick in the foundation of the construction \eqref{quantAproc} based on the topological recursion,
which after a few more systematic and completely rigorous steps builds the $q$-difference operator as a power series in~$\hbar$:
\beq
A(x,y) \quad \leadsto \quad \op A(\op x, \op y; q) = A(\op x, \op y ) + \hbar A_1(\op x, \op y) + \cdots \,.
\label{quantAprocpert}
\eeq
Even though we omit the intermediate steps due to constraints of space, the reader should simply be aware
that a well-defined, systematic procedure exists. The existence and uniqueness of this procedure are well-motivated based on physical considerations; in fact, these form one of the basic premises of quantum mechanics.

By looking at \eqref{quantAprocpert} it would seem that we would therefore have to compute terms to arbitrarily high order in this series to write down the operator~$\op A$. However, in practice, this is not the case; we usually need to compute only one or sometimes two terms in the series to know $\op A$ exactly! The trick is as follows: if we know, \textsl{a priori}, that the operator we construct can be written as a rational function of~$q=e^\hbar$, then the higher order terms in the expansion in~$\hbar$ must resum nicely into an expression of this form. We also have information about the classical limit ($q\goesto 1$) of this expression.  Armed with this information, it is usually pretty straightforward to construct the quantization of~$A(x,y)$ in closed form.

For example, if we know both the classical term and the first quantum correction $\hbar A_1(\op x, \op y)$
in the expansion \eqref{quantAprocpert}, there is a good chance we can reconstruct the quantum operator
\beq
\op A (\op x, \op y; q) \; = \; \sum_{m,n}\, a_{m,n}\, q^{c_{m,n}}\, \op x^m\, \op y^n
\label{Aqnice}
\eeq
simply from the data $\{ a_{m,n} \}$ of the original polynomial $A(x,y) = \sum a_{m,n} x^m y^n$
and from the exponents $\{ c_{m,n} \}$ determined by $\hbar A_1(\op x, \op y)$.
This trick becomes especially useful for curves that come from knots and 3-manifolds.
Indeed, in such examples the leading quantum correction is determined by
the ``classical'' knot invariant $\Delta (q)$ called the twisted Alexander polynomial.
Therefore, a simple mnemonic rule to remember what goes into the construction
of the operator $\op A (\op x, \op y; q)$ in such situations can be schematically expressed as \cite{abmodel}:
\beq
\boxed{\phantom{\int}
`` \; A(x,y) + \Delta (q) \; \Rightarrow \; \op A(\op x, \op y; q) \; "
\phantom{\int}}
\eeq
Concretely, the exponents $c_{m,n}$ in \eqref{Aqnice} can be determined by requiring that the relation
\beq
2 \sum_{m,n}\, a_{m,n}\, c_{m,n}\, x^m y^n \; = \;
\frac{\partial A}{\partial \ln x} \left( \frac{\partial A}{\partial \ln y} \right)^{-1} \frac{\partial^2 A}{(\partial \ln y)^2}
+ x \frac{\partial \Delta(x)}{\partial x} \frac{\partial A}{\partial \ln y}
\eeq
holds for all values of $x$ and $y$ (along with $A(x,y)=0$). 

\begin{example}	\label{ex:q-Apol-trefoil}
Consider once more the trefoil knot $K=3_1$, which has $A$-polynomial $A(x,y)= (y^{-1}-1)(y+x^3)$ and where,
following our earlier agreement, we replaced $x^2$ by $x$ to simplify the expressions, cf. \eqref{eq:A-trefoil1}.
Notice, that $A(x,y)$ in this example is a degree-2 polynomial in $y$.
Quantization \eqref{quantAprocpert} then gives an operator which is also of degree 2 in $\op y$
\beq
\op A(\op x, \op y; q) = \alpha \op y^{-1} + \beta + \gamma \op y,
\label{Aq-31}
\eeq
where
\beq
\begin{cases}
\alpha = \frac{x^2(x-q)}{x^2-q}; &\\
\beta = q\left( 1 +x^{-1} -x + \frac{q-x}{x^2 - q} - \frac{x-1}{x^2 q - 1} \right); &\\
\gamma = \frac{q-x^{-1}}{1-qx^2}. &
\end{cases}
\eeq
In the representation \eqref{xnyshift}, our quantized constraint \eqref{op-zero}
then gives an operator relation that takes the form of a recurrence in the variable~$n$:
\beq
\op A Z = 0 \implies \alpha(q^n;q) Z_{n-1} + \beta(q^n;q) Z_n + \gamma(q^n;q) Z_{n+1} = 0,
\label{31recurrence}
\eeq
where we recall that~$n$ was defined  so that $x=q^n$.
\end{example}

\begin{exercise}	\label{ex:rec-Apol-trefoil}
Solve this recurrence with the initial conditions
\beq
Z_n = 0\text{ for } n\leq 0; \quad Z_1=1. \eeq
 That is, find the first few terms of the  sequence $Z_n(q)$ for $n=2,3,\ldots$
 \begin{proof}[Solution]
Straightforward computation gives
\beq
\begin{aligned}
Z_2(q) &= -\beta(q;q)/\gamma(q;q) \\
&= -\frac{1-q^3}{q-q^{-1}}\cdot q\left(1+q^{-1}-q-\frac{q-1}{q^3-1} \right) \\
&= - \frac{(1-q^3)(1+q-q^2) + q(q-1)}{q-q^{-1}}\\
&=  \frac{-1 + q^3 + q^4 - q^5 }{q-q^{-1}} \\
&=  q + q^3 - q^4,
\end{aligned}
\eeq
as well as
\beq
\begin{aligned}
Z_3(q) &= - (\alpha(q^2;q) + \beta(q^2;q) Z_2(q))/\gamma(q^2;q) \\
&= q^2 + q^5 - q^7 + q^8 - q^9 - q^{10} + q^{11},\\
\end{aligned}
\eeq
after a little manipulation. Notice that the~$Z_n$ all turn out to be polynomials!
\end{proof}
\end{exercise}

Now, we come to one of the punch lines of these lectures. The reader who has completed Exercise~\ref{hw:HOMFLY} and followed through the derivation of \eqref{J-trefoil} may have noticed a startling coincidence: $Z_n$ produced by our our recurrence relation~\eqref{31recurrence} is none other than the $n$-colored Jones polynomial; that is, the quantum group invariant of the knot decorated with extra data consisting of the Lie algebra $\lie{g} = \lie{sl}(2)$ and its $n$-dimensional representation $R=V_n$.

This is no coincidence, of course. As we reviewed in section \ref{sec:intro}, the $n$-colored Jones polynomial is simply the partition function of Chern-Simons TQFT with gauge group $G=SU(2)$. On the other hand, in this section we explained that the classical $A$-polynomial and its quantum, non-commutative version have a natural home in Chern-Simons TQFT with complex gauge group $G_{\C}=\SL(2,\C)$. In particular, we saw how the usual rules of quantum mechanics replace the classical constraint~\eqref{ACurve} with an operator relation \eqref{op-zero},
\beq
\mathscr{C} : A(x,y) = 0 \quad \leadsto \quad
\op A(\op x, \op y; q) Z_\text{CS}(M) = 0 \,,	\label{qdiff}
\eeq
where $Z_\text{CS}(M)$ is the state vector associated by quantization to the Lagrangian submanifold~$\mathscr{C}$ (or, equivalently, associated by the Chern-Simons TQFT functor to the 3-manifold~$M$).
Since $G_{\C}=\SL(2,\C)$ is a complexification of $G=SU(2)$, the partition functions in these two theories are closely related \cite{DGLZ,Wit-anal}.
In particular, it was argued in \cite{gukov} that \textsl{both}~$\SU(2)$ and~$\SL(2,\C)$ partition functions must satisfy the quantum constraint equation~\eqref{qdiff}. In the $n$-representation \eqref{xnyshift} it takes the form of a \vocab{recurrence relation}
\beq
A(x,y) = \sum_k a_k(x) y^k \quad \leadsto \quad
\sum_k  a_k (q^n;q) J_{n+k} (K;q) = 0 \,,
\label{recurrence}
\eeq
which is precisely our $q$-difference equation \eqref{31recurrence} in the above example, where~$K$ was taken to be the trefoil knot.
More generally, the equation~\eqref{recurrence} is a $q$-difference equation, describing the behavior with respect to $n$, or ``color dependence,'' of the $n$-colored Jones polynomial that is computed by Wilson loop operators in the $\SU(2)$ Chern-Simons theory.

The relation between the quantization of the $A$-polynomial and the quantum group invariants \eqref{recurrence}
that follows from Chern-Simons theory is the statement of the \vocab{quantum volume conjecture} \cite{gukov}
(see \cite{dimofte} for a review of earlier developments that led to it).
This conjecture was independently proposed in \cite{garoufalidis} around the same time and is also know as the \vocab{AJ-conjecture}.
It provides a bridge between two seemingly distant areas of knot theory, the classical $A$-polynomial and the study of quantum group invariants. Before the discovery of this connection, the separate communities of knot theorists working on these two different types of invariants had very little contact with one another.

Do two knots having the same $A$-polynomial always have all the same $n$-colored Jones polynomials? Based on the above connection, we would expect an affirmative answer, given that the quantization procedure for the $A$-polynomial is essentially unique. This has been checked for knots up to large number of crossings, although there is as yet no formal proof. If it is true, then a single algebraic curve constructed without any reference to quantum groups encodes all the information about the whole tower of $n$-colored Jones polynomials:
\beq
A(x,y)
\quad \leadsto \quad
\op A(\op x, \op y; q)
\quad \leadsto \quad
J_{n} (K;q) \,.	\label{AAJ}
\eeq
Nonetheless, even if all the $n$-colored Jones polynomials together carry no more information than the $A$-polynomial, their relation to quantum groups still makes them interesting objects of study in their own right. (It is also worth noting that the study of the colored Jones polynomial predates the discovery of the $A$-polynomial.)

Once we explained how to go, via quantization, from the classical $A$-polynomial to quantum group invariants \eqref{AAJ}
it is natural to ask whether there is a simple way to go back. The \vocab{generalized volume conjecture} \cite{gukov}
proposes an affirmative answer to this question and is also based on the fact that the analytic continuation of~$\SU(2)$ is~$\SL(2,\C)$.
It states that, in the classical limit $q \to 1$ accompanied by the ``large color'' limit $n \to \infty$,
the $n$-colored Jones polynomial, as a Wilson line in $\SU(2)$ Chern-Simons theory \cite{WittenCS}, exhibits the exponential behavior
\beq
J_n (K; q = e^{\hbar}) \mathrel{ \mathop{\mathop{\sim}_{n\goesto\infty}}_{\hbar\goesto 0} } \exp\left( \frac{1}{\hbar} S_0(x) + \cdots \right),
\label{genVC}
\eeq
where the limits are taken with $q^n = x$ held fixed. Here $S_0(x)$ is the classical action of $\SL(2,\C)$ Chern-Simons theory, which is
\beq
S_0(x) = \int \log y \frac{dx}{x}
\label{classical-action}
\eeq
evaluated on a path within the curve $\mathscr{C}:A(x,y) = 0$. Here, by an abuse of notation, the variable $x$ stands in for a point on the Riemann surface; $S_0$ is actually a function on~$\mathscr{C}$, and the integral in~\eqref{classical-action} is taken along a path in~$\mathscr{C}$ from some fixed base point to the point at which $S_0$ is evaluated. Moreover, \eqref{classical-action} is only well defined if the integrality condition~\eqref{integrality} holds! The change $\Delta S_0$ that comes from composing the path used in our evaluation with an arbitrary closed cycle must be valued in~$ 2\pi \Z$, so that the quantity $e^{iS_0}$ is well-defined and independent of path; the integrality condition ensures that this is so.

To summarize, the generalized volume conjecture gives us two important ways of thinking about the $A$-polynomial: firstly, as a characteristic variety encoding information about $\SL(2,\C)$ flat connections, and secondly, as a \vocab{limit shape} in the limit of large color.

We have now begun to see how the seemingly disparate topics we have been discussing are connected to one another. Roughly speaking, there are four  major themes in these lectures: quantum group invariants, the $A$-polynomial, quantization, and categorification. We have now seen how quantization relates the $A$-polynomial and quantum group invariants, providing a bridge between seemingly unrelated knot polynomials. In what remains, we will return to ideas of categorification, hoping to give at least a glimpse of how knot polynomials arise from deeper and more powerful homological invariants.

\section{Categorification}
\label{sec:categorification}
\index{categorification}

Categorification is a powerful and flexible idea; it can mean different things in different contexts, and a given mathematical construction may admit more than one categorification depending on how one chooses to look at its structure. In the context of topological quantum field theories, however, categorification is manifested in a very natural way. The categorification of a 3-dimensional TQFT should be a 4-dimensional TQFT, from which the 3D theory is recovered by dimensional reduction, see e.g. \cite{CFrenkel,gukovRTN}. That is,
\[
\xymatrixcolsep{1.75in}
\xymatrix{
\text{3D TQFT} \ar@/^/[r]^{\text{categorification}} & \text{4D TQFT} \ar@/^/[l]^{\text{dimensional reduction}}
}
\]
We can tabulate the information that each of these TQFTs should associate to geometrical objects in the below table:
\smallskip \begin{center}
\begin{tabular}{|l|l|l|} \hline
\textsl{Geometry} & \textsl{3D TQFT} & \textsl{4D TQFT} \\ \hline
3-manifold $M$, & number $Z(M)$, & vector space $\Hil_K$ \\
knot $K\subset M$ & polynomial invariant $P(K)$ & \\ \hline
2-manifold $\Surf$ & vector space $\Hil_\Surf$ & category $\text{\bf Cat}_\Surf$ \\ \hline
\end{tabular}
\end{center} \smallskip
Thus, to a geometrical object of given dimension, a categorified TQFT associates objects of one higher categorical level than its decategorified counterpart. (The categorical level of the object associated by a TQFT to something in geometry corresponds to its \textsl{codimension}, so that a 4D TQFT assigns numerical invariants to 4-manifolds. Famous examples of these are given by Donaldson theory \cite{WittenDonaldson} and Seiberg-Witten theory \cite{WittenMonopoles}.)

In 2000, Mikhail Khovanov~\cite{khovanov} succeeded in constructing a categorification of the Jones polynomial. Like the Jones polynomial, it is associated to the extra data $\lie{g} = \lie{sl}(2)$ and its fundamental representation $R=V_2$. To give the barest outline, his construction associates a chain complex to a diagram of a link~$K$. The homology of this chain complex can be shown to be invariant under the Reidemeister moves, and therefore to be an invariant of~$K$. \vocab{Khovanov homology}~$\Khom(K)$ is doubly graded, and the Jones polynomial is its graded Euler characteristic, cf. \eqref{Eulerhom},
\beq
J(q) = \sum_{i,j} (-1)^i q^j \dim \Khom(K) \,.
\label{JfromKh}
\eeq
Sometimes it is convenient to encode information about the Khovanov homology in its Poincar\'e polynomial:
\beq
\Kh(q,t) = P_{\lie{sl}(2),V_2} (q,t) = \sum_{i,j} t^i q^j \dim \Khom(K).
\eeq
The Jones polynomial is then recovered by making the evaluation at $t=-1$. As an example, the Poincar\'e polynomial of the trefoil knot is
\beq
\Kh(q,t; K=3_1) = q + q^3 t^2 + q^4 t^3 \,.
\label{Kh31}
\eeq
It is easy to see that the evaluation at $t=-1$ indeed returns the normalized Jones polynomial of the trefoil knot \eqref{J-trefoil}
that we saw in section \ref{sec:intro}. By definition, this version of the homology is called \vocab{reduced}.
Its close cousin, the \vocab{unreduced knot homology} categorifies the \vocab{unnormalized} polynomial invariant.
Thus, for the unnormalized Jones polynomial \eqref{trefJones} of $K=3_1$ the corresponding categorification
is given by the unreduced Khovanov homology shown in Figure \ref{trefoil-khovanov}.

\begin{figure}
\begin{tabular}{c|ccccccccc}
\llap{$i = {}$}3 & $\cdot$&$\cdot$ &$\cdot$ &$\cdot$ &$\cdot$ &$\cdot$ &$\cdot$ &$\cdot$ & $\Z$ \\
2 &$\cdot$&$\cdot$&$\cdot$&$\cdot$& $\Z$ &$\cdot$& $\Z/2$ &$\cdot$&$\cdot$ \\
1 &$\cdot$&$\cdot$&$\cdot$&$\cdot$&$\cdot$&$\cdot$&$\cdot$&$\cdot$&$\cdot$ \\
0 & $\Z$ &$\cdot$& $\Z$ &$\cdot$&$\cdot$&$\cdot$&$\cdot$&$\cdot$&$\cdot$ \\ \hline
& 1& 2& 3& 4& 5& 6& 7& 8& 9\rlap{${}=j$}
\end{tabular}
\caption{The Khovanov homology $\Khom(K=3_1)$ of the trefoil knot.
\label{trefoil-khovanov}}
\end{figure}

Much like the Khovanov homology of a knot is a categorification of its Jones polynomial or quantum~$\lie{sl}(2)$ invariant,
there exist generalizations \cite{Yonezawa,Wu,Webster,CKrushkal,Stroppel}
of the Khovanov homology categorifying the $n$-colored Jones polynomials for all~$n$:
\beq
J_n(K;q) = P_n(K;q,t)|_{t=-1} = \left. \sum_{i,j} q^i t^j \dim H^{\lie{sl}(2),V_n}_{i,j} (K) \right|_{t=-1}.
\label{JnPn}
\eeq
The $n$-colored $\lie{sl}(2)$ knot homologies satisfy recursion relations, just like their decategorified versions, and exhibit beautiful asymptotic behavior in the limit of large~$n$. Both of these behaviors are controlled by a \vocab{refined algebraic curve}, which is an analogue of the $A$-polynomial~\cite{FGS}:
\beq
\mathscr{C}^\text{ref}:~ A^\text{ref}(x,y;t) = 0 \,.	\label{apol-t}
\eeq
This curve is a $t$-deformation of (the image of) the representation variety of a knot complement~$M$ in the classical phase space of the Chern-Simons theory, which is the moduli space $\moduli_\text{flat}(\SL(2,\C), \Surf)$ of flat connections. Here $\Surf=\bdy M$. Much like the representation variety \eqref{ACurve} of~$M$, its $t$-deformation \eqref{apol-t} is a holomorphic Lagrangian submanifold with respect to the symplectic form \eqref{xysympform}.

\begin{example}
In section \ref{sec:Apol-1} we derived the $A$-polynomial of the trefoil knot \eqref{eq:A-trefoil1}.
Then, in section \ref{sec:quantization} we discussed its quantization, or \vocab{non-commutative} $q$-deformation.
In both cases, the result is a quadratic polynomial in $y$. Similarly, the \vocab{commutative} $t$-deformation
of the $A$-polynomial for the trefoil knot is a quadratic polynomial in $y$,
\beq
A^\text{ref}(x,y;t) = y^2 - \frac{1-xt^2+x^3 t^5 + x^4 t^6 + 2x^2 t^2 (t+1)}{1+xt^3} y + \frac{(x-1)x^3t^4}{1+xt^3}
\label{A-ref31}
\eeq
which reduces to the ordinary $A$-polynomial \eqref{eq:A-trefoil1} in the limit $t=-1$.
\end{example}

As in section~\ref{sec:quantization}, quantization of $\moduli_\text{flat}(\SL(2,\C), \Surf)$ with its natural symplectic form promotes $x$ and~$y$ to operators obeying the commutation relation
\beq
\op y \op x = q \op x\op y
\eeq
and turns the planar algebraic curve \eqref{apol-t} into a $q$-difference recursion relation, cf. \eqref{recurrence},
\beq
\boxed{\phantom{\int}
\op A^\text{ref} P_\star (K;q,t) \simeq 0 \,,
\phantom{\int}}
\label{APconj}
\eeq
where $\op x P_n = q^n P_n$ and $\op y P_n = P_{n+1}$. This recursion relation, called the \vocab{homological volume conjecture} in \cite{FGS}, provides a natural categorification of the generalized volume conjecture that was the subject of section \ref{sec:quantization}. Unlike the generalized volume conjecture, its homological version \eqref{APconj} is based on a much more sophisticated physics that involves a physical interpretation of knot homologies in terms of refined BPS invariants \cite{GSV,gukovRTN} and dynamics of supersymmetric gauge theories \cite{DGH,fiveknots,DGGindex,FGSS}. The details of this physical framework go way beyond the scope of these lectures and we simply refer the interested reader to the original papers.\\


There also exists a homology theory categorifying the HOMFLY-PT polynomial~\cite{KhR1,KhR2}. As should be obvious, this theory must be triply graded; the HOMFLY-PT polynomial is recovered by taking the graded Euler characteristic, cf. \eqref{JfromKh},
\beq
P_{a,q}(K) = \sum_{ijk} (-1)^i q^j a^k \dim\mathscr{H}_{ijk}(K) \,.
\label{PfromH}
\eeq
Just as we did for \vocab{Khovanov homology}, we can construct the Poincar\'e polynomial associated to the \vocab{HOMFLY homology}, which will encode information about the dimensions of its groups at each level:
\beq
\mathscr{P}(a,q,t) = \sum_{ijk} t^i q^j a^k \dim \mathscr{H}_{ijk}(K) \,.
\eeq
Then decategorification corresponds once more to evaluation at the value~$t=-1$. It turns out that even the HOMFLY homology is not a complete invariant of knots; nonetheless, these homological invariants are strictly finer and stronger than their decategorified counterparts.
For instance, HOMFLY homology can distinguish between the knots $5_1$ and~$10_{132}$, discussed earlier, that have identical Jones, Alexander, and HOMFLY-PT polynomials \eqref{PP5110132}.

We should remark also that $n$-colored generalizations of HOMFLY homology can be constructed, and that the color dependence can be encoded in an algebraic curve, just as the zero locus of the $A$-polynomial encodes the information about color dependence of the $n$-colored Jones polynomial. We will return to this point and discuss the corresponding algebraic curve in much more detail in the final section of these lectures. Meanwhile, in the rest of this section we mostly focus on the ordinary, \textsl{uncolored} HOMFLY homology aiming to explain its structure and how to compute it in practice.

As we shall see, the structure of the homological knot invariants turns out to be so rich and so powerful that, once we learn enough about it, we will be able to compute, say, the Khovanov homology and the HOMFLY homology of the trefoil knot solely from the data of its Jones polynomial. In other words, in a moment we will learn powerful techniques that will allow us to reproduce \eqref{Kh31} without even learning the definition of Khovanov homology. And, much of this structure is present --- in fact, in a richer form! --- in the \vocab{colored HOMFLY homology} as well \cite{GS}.

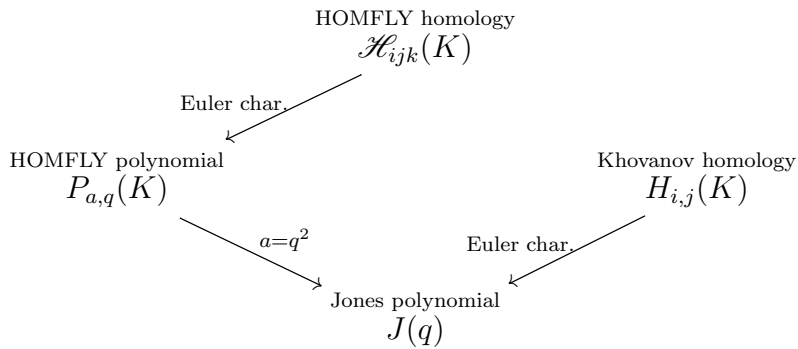
\begin{figure}
\[
\xymatrix{
& \stackrel{\text{HOMFLY homology}}{\mathscr{H}_{ijk}(K)} \ar[dl]_{\text{~~~Euler char.~~~}} & \\
\stackrel{\text{HOMFLY polynomial}}{P_{a,q}(K)} \strut \ar[dr]^{a=q^2} & &  \stackrel{\text{Khovanov homology}}{\Khom(K)} \ar[dl]_{\text{~~~Euler char.~~~}} \\
& \stackrel{\text{Jones polynomial}}{J(q)} &
}
\]
\caption{A summary of relations between homological and polynomial invariants.}
\label{hom/poly}
\end{figure}

Let us start by summarizing the familiar relations \eqref{JonesPq2}, \eqref{JfromKh}, \eqref{PfromH}
between homological and polynomial invariants diagramatically, as shown in Figure~\ref{hom/poly}.
We would like to be able to fill in the missing fourth arrow, i.e., to have a way of recovering Khovanov homology directly from the HOMFLY homology. This, however, is rather delicate for a number of reasons. First, the specialization $a=q^2$ does not make sense in the context of the homology theories. At best one could try to complete the diagram by working with the Poincar\'e polynomials associated to these theories:
\beq
\xymatrixrowsep{0.15in}
\xymatrix{
& \mathscr{P}(a,q,t) \ar[dl]_{t=-1} \ar[dr]^{a=q^2} & \\
P(a,q) \ar[dr]^{a=q^2} & & \Kh (q,t) \ar[dl]_{t=-1} \\
& J(q) &
} \label{poincare-diagram}
\eeq
As we explain shortly, even this is too naive due to a simple, yet conceptual reason.
Nevertheless, for a moment let us ignore this issue and proceed as if \eqref{poincare-diagram} were actually correct.

\begin{example}
Let us see if we can use the information in~\eqref{poincare-diagram} to reconstruct~$\mathscr{P}(a,q,t)$ for the trefoil knot.
We know already that
\beq
\begin{cases}
P(a,q) = aq^{-1} + a q - a^2, & \\
\Kh(q,t) = q + q^3 t^2 + q^4 t^3. &
\end{cases} \eeq
We can attempt to guess $\mathscr{P}(a,q,t)$ just by comparing terms; this gives
\beq
\mathscr{P}(a,q,t) = a q^{-1} + a q t^2 + a^2 t^3.		\label{trefoil-aqt}
\eeq
This naive guess turns out to be correct! Using only information from the HOMFLY-PT polynomial and Khovanov homology (both of which are easily computable), we have obtained information about the triply-graded HOMFLY homology theory, which encodes information about the~$\lie{sl}(N)$ homological invariants for all~$N$.
\end{example}

In fact, one can even get to~\eqref{trefoil-aqt} without knowing the Khovanov homology! Our task is to assign a $t$-degree to each term in the HOMFLY-PT polynomial. We can do this using the following trick: From Exercise~\ref{hw:HOMFLY}, the reader should know that evaluating $P(a,q)$ at $a=q$ yields a monomial (exactly which monomial depends on a simple knot invariant and a choice of normalization). This turns out to be true for any knot: the HOMFLY-PT polynomial will always become trivial, i.e., monomial, when evaluated at $a=q$.
Therefore, to ensure the needed cancellation when the specialization $a=q$ is made,
the normalized HOMFLY-PT polynomial for any knot must have the following schematic form:
\beq
P_{a,q} = 1 + (1-a^{-1}q) Q(a,q),
\eeq
where~$Q$ is some polynomial factor.
The basic reason for this is that taking $a=q$ corresponds to asking about the $\lie{sl}(1)$ polynomial invariant, which must always be trivial. A similar simplification happens in the case $a=q^{-1}$.

What about the $\lie{sl}(1)$ \textsl{homological} invariant? Since $\mathscr{P}(a,q,t)$ has only positive coefficients, $\mathscr{P}(q,q,t)$ can't be trivial --- it must reduce to a monomial only because of cancellations that occur for $t=-1$. But we would not expect to be able to construct any nontrivial invariants with $\lie{sl}(1)$, homological or otherwise. This is a clue that something more sophisticated must be happening in the way that one extracts Khovanov homology (generally, $\lie{sl}(N)$ homology) from the HOMFLY homology.

The reason, to which we alluded earlier, is that when polynomial knot invariants are categorified
one correspondingly needs to upgrade the specialization $a=q^N$ of section \ref{sec:intro} to homological level.
In other words, trying to use the specialization $a=q^N$ as we did in diagram \eqref{poincare-diagram}
is too naive and the suitable operation should also be from the world of homological algebra.

It turns out that the correct homological lift of the specialization $a=q^N$ involves a conceptually new ingredient,
which has no analog at the (decategorified) polynomial level: a family of differentials $\{ d_N \}$ on the HOMFLY homology, indexed by $N\in\Z$.
These differentials endow HOMFLY homology with a structure that is much richer than what can be seen at the polynomial level
and that is responsible for our claim that \eqref{trefoil-aqt} can be derived even without the knowledge of the Khovanov homology.
By viewing the triply-graded homology as a complex and taking its homology with respect to this differential, one recovers the doubly-graded Khovanov homology. Specifically, in the grading conventions of \cite{GS}, the differentials have degree
\beq
\begin{aligned}
d_{N > 0} : & (-1,N,-1), \\
d_{N \leq 0} : & (-1,N,-3) \end{aligned}
\label{dNdegree}
\eeq
with respect to~$(a,q,t)$ grading. The homology of $\mathscr{H}_\star$, viewed as a complex with differential~$d_N$, returns the doubly-graded $\lie{sl}(|N|)$ homology theory \cite{KhR1} or the \vocab{knot Floer homology} \cite{OShfk,RasmussenHFK} in the special case $N=0$, see \cite{DGR} for details. In particular, its homology with respect to the differentials~$d_1$ and~$d_{-1}$ must be trivial.

For instance, in considering the reduction of HOMFLY homology to the $\lie{sl}(1)$ homological invariant, almost all of the terms in the triply-graded HOMFLY homology will be killed by the differential $d_1$, leaving behind a ``trivial'' one-dimensional space,
\beq
\dim \left( \mathscr{H}_\star , d_1 \right) = 1 \,.
\eeq
Because the differential $d_1$ has definite grading \eqref{dNdegree}, the Poincar\'e polynomial of HOMFLY homology therefore must be of the following general form
\beq
\mathscr{P}(a,q,t) = 1 + (1+a^{-1} q t^{-1} ) Q_+ (a,q,t) \,,
\label{d1diff}
\eeq
where the first term represents a contribution of the (trivial)~$\lie{sl}(1)$ knot homology, and $Q_+ (a,q,t)$ is some polynomial
with positive coefficients. Note, that the Poincar\'e polynomial \eqref{d1diff} necessarily has all of its coefficients nonnegative.
Similar structure follows from the existence of another \vocab{canceling differential} $d_{-1}$ that also kills all but one generators of the HOMFLY homology. The physical interpretation of the differentials $\{ d_N \}$ can be found in \cite{GS}.

Now, just from the little we learned about the differentials $d_1$ and $d_{-1}$, we can reconstruct the HOMFLY homology of the trefoil knot. First, we can get information about the $a$- and $q$-degrees of nontrivial HOMFLY homology groups just from the HOMFLY-PT polynomial. For the trefoil knot, these are depicted below:
\begin{center}
\begin{tikzpicture}
\draw (-1,0.5) node[anchor=north]{$-1$};
\draw (0,0.5) node[anchor=north]{$0$};
\draw (1,0.5) node[anchor=north]{$1$};
\draw (-2,1) node[anchor=east]{$1$};
\draw (-2,2) node[anchor=east]{$2$};
\draw[->] (-2.5,0.5) -- (1.5,0.5) node[anchor=west]{$q$};
\draw[->] (-2,0) -- (-2,2.5) node[anchor=east]{$a$};
\draw (-1,1) node[fill,circle,scale=0.65]{};
\draw (0,2) node[fill,circle,scale=0.65]{};
\draw (1,1) node[fill,circle,scale=0.65]{};
\draw[dashed,->] (0,2) -- (-0.8,1.2);
\draw (-0.5,1.5) node[anchor=south east]{$d_{-1}$};
\draw[dashed,->] (0,2) -- (0.8,1.2);
\draw (0.5,1.5) node[anchor=south west]{$d_{1}$};
\end{tikzpicture}
\end{center}
It is clear that each of the differentials $d_{\pm 1}$ can only act nontrivially in one place. From the condition that they give rise to trivial homology, each must be surjective; this determines the relative $t$-degree of each group. Taking the point with $(a,q)$-degree $(1,-1)$ to have~$t=0$, it immediately follows that the degrees of the other groups with respect to $(a,q,t)$ degree are $(2,0,3)$ and $(1,1,2)$. We have now managed to extract this information without even computing Khovanov homology; the results of Exercise~\ref{hw:HOMFLY} and the above trick are all we need.

\section{Epilogue: super-$A$-polynomial}	\label{sec:Apol-2}

In this section, we give a somewhat deeper discussion of the connection between physics, homological knot invariants, and the quantization of the $A$-polynomial, constructing one final bridge between the ideas of quantization and categorification. This final section of the lectures can be seen as an addendum; based on recent progress \cite{AVqdef,superA,Nawata,FGSS} it summarizes material that was covered in a talk given at the conference following the summer school, and so is somewhat more technical.

In these lectures, we saw several deformations of the classical $A$-polynomial $A(x,y)$ introduced in section \ref{sec:Apol-1}.
Thus, in section \ref{sec:quantization} we saw how quantization of $\SL(2,\C)$ Chern-Simons theory leads to a \vocab{non-commutative} $q$-deformation \eqref{Aquantproc}. Then, in section \ref{sec:categorification} we saw how more sophisticated physics based on refined BPS invariants leads to a categorification of the generalized volume conjecture and a \vocab{commutative} $t$-deformation \eqref{apol-t}.

These turn out to be special cases of a more general three-parameter ``super-deformation'' of the $A$-polynomial introduced in \cite{superA}. Two out of these three deformations are \vocab{commutative} and will be parametrized by $a$ and $t$, while the third \vocab{non-commutative} deformation is produced essentially by the quantization procedure \eqref{quantAprocpert} of section \ref{sec:quantization}:
\beq
A^\text{super} (x,y;a,t) \quad \leadsto \quad \op A^\text{super} (\op x, \op y; a,q,t) \,.
\label{superAquant}
\eeq
What is the meaning of this \vocab{super-$A$-polynomial}?

The best way to answer this question is to consider an example. In fact, let us repeat the analogs of
Example~\ref{ex:q-Apol-trefoil} and Exercise~\ref{ex:rec-Apol-trefoil}:

\begin{example}
For our favorite example, the trefoil knot $K=3_1$,
we know from our earlier discussion that the classical $A$-polynomial $A(x,y) = (y-1)(y+x^3)$ is quadratic in $y$,
and so are its $t$-deformation \eqref{A-ref31} and $q$-deformation \eqref{Aq-31}.
The same is true of the super-$A$-polynomial of $K=3_1$,
\beq
\begin{split}
A^\text{super}(x,y;a,t) =
y^2 - \frac{a \left( 1-t^2x + 2t^2 (1+at) x^2 + a t^5 x^3 + a^2 t^6 x^4 \right) }{1+at^3x} y\\
 + \frac{a^2 t^4 (x-1) x^3}{1+at^3 x}.
 \end{split}	\label{A-super}
\eeq
which clearly reduces to \eqref{A-ref31} upon setting $a=1$ and to the ordinary $A$-polynomial \eqref{eq:A-trefoil1}
upon further specialization to $t=-1$. Moreover, the quantization procedure of section~\ref{sec:quantization}
turns super-$A$-polynomial \eqref{A-super} into a $q$-difference operator, which can be interpreted as a recurrence relation, similar to \eqref{superAquant},
\begin{multline}
\op A^\text{super}(\op x, \op y; a, q, t) = \alpha + \beta \op y + \gamma \op y^2 \\
\implies \alpha \mathscr{P}_n + \beta \mathscr{P}_{n+1} + \gamma \mathscr{P}_{n+2} = 0 \,.
\label{31superrecurrence}
\end{multline}
Here, the coefficients $\alpha$, $\beta$, and~$\gamma$ are certain rational functions of the variables $a$, $q$, $x \equiv q^n$, and~$t$, whose explicit form can be found in \cite{superA}.
\end{example}

\begin{exercise}
As in Exercise \ref{ex:rec-Apol-trefoil}, solve the recurrence \eqref{31superrecurrence} with the initial conditions
\beq
\mathscr{P}_n = 0\text{ for } n \leq 0; \quad \mathscr{P}_1=1. \eeq
That is, find the first few terms of the sequence $\mathscr{P}_n(q)$ for $n=2,3,\ldots$
\begin{proof}[Solution]
Straightforward computation gives:
\begin{center} {\small
\begin{tabular}{c|l}
$n$ & $\mathscr{P}_n(a,q,t)$ \\ \hline
1 & 1 \\
2 & $aq^{-1} + aqt^2 + a^2 t^3$ \\
3 & $ a^2 q^{-2} + a^2 q(1+q) t^2 + a^3 (1+q) t^3 + a^2 q^4 t^4 + a^3 q^3 (1+q) t^5 + a^4 q^3 t^6$ \\
4 & $a^3 q^{-3} + a^3 q (1+q+q^2) t^2 + a^4 (1+q+q^2) t^3 + a^3 q^5 (1+q+q^2) t^4 + {}$ \\
& ${}+a^4 q^4 (1+q) (1+q+q^2) t^5 + a^3 q^4 (a^2 + a^2 q + a^2 q^2 + q^5) t^6 + {}$ \\
& ${}+ a^4 q^8 (1+q+q^2) t^7 + a^5 q^8 (1+q+q^2) t^8 + a^6 q^9 t^9$ \\
\end{tabular} }
\end{center}
How should we interpret these polynomial invariants? The answer can be guessed from a couple of clues in the above table: firstly, all $\mathscr{P}_n(a,q,t)$ involve only positive integer coefficients. Secondly, we have seen $\mathscr{P}_2(a,q,t)$ before; it is the Poincar\'e polynomial \eqref{trefoil-aqt} of the triply-graded HOMFLY homology of the trefoil knot!
\end{proof}
\end{exercise}

These considerations lead one to guess, correctly, that $\mathscr{P}_n(a,q,t)$ is the Poincar\'e polynomial of the $n$-colored generalization of the HOMFLY homology:
\beq
\mathscr{P}_n(a,q,t) = \sum_{ijk} t^i q^j a^k \dim \mathscr{H}_{ijk}^{(n)} (K) \,.
\eeq
Naively, one might expect that the specialization $a=q^2$ in the polynomial $\mathscr{P}_n (a,q,t)$
should return the $n$-colored $\lie{sl}(2)$ homology in \eqref{JnPn}, and so forth.
However, in the homological world, this specialization is a little bit more subtle.
It turns out that, just as  we saw earlier in section \ref{sec:categorification},
the colored homology~$\smash{\mathscr{H}_{ijk}^{(n)} (K)}$ comes naturally equipped with a family of differentials~$d_N$; viewing~$\smash{\mathscr{H}_{ijk}^{(n)} (K)}$ as a complex and taking its homology with respect to the differential~$d_2$
allows one to pass directly from the $n$-colored HOMFLY homology to the $n$-colored analog of the Khovanov homology.

To summarize, the super-$A$-polynomial encodes the ``color dependence'' of the \vocab{colored HOMFLY homology},
much like the ordinary $A$-polynomial and its \vocab{$t$-deformation} do for the colored Jones polynomial \eqref{recurrence}
and the colored $\lie{sl}(2)$ homology \eqref{APconj}, respectively:
\beq
\boxed{\phantom{\int}
\op A^\text{super} \mathscr{P}_\star (a,q,t) \simeq 0 \,.
\phantom{\int}}
\label{superAPconj}
\eeq
Moreover, setting $q=1$ gives the classical super-$A$-polynomial with two commutative parameters $a$ and $t$.
Its zero locus defines an algebraic curve
\beq
\mathscr{C}^\text{super}:~ A^\text{super}(x,y;a,t) = 0 \,.	\label{apol-at}
\eeq
which in various limits reduces to the $A$-polynomial curve \eqref{rep-variety} and its ``refined'' version \eqref{apol-t}.
This curve plays the same role for colored HOMFLY homology as the ordinary $A$-polynomial does for the colored Jones invariants.
Specifically, there is an obvious analog of the generalized volume conjecture \eqref{genVC},
which states that \eqref{apol-at} is the \vocab{limit shape} for the $S^n$-colored HOMFLY homology
in the large color limit $n \to \infty$ accompanied by $q \to 1$ \cite{superA}.

A simple way to remember different specializations of the two-parameter ``super-deformation'' of the $A$-polynomial
is via the following diagram:
\beq
\xymatrixcolsep{0.2in}
\xymatrix{
& {A^\text{super} (x,y;a,t)} \ar[dl]_{a=1} \ar[dr]^{t=-1} & \\
A^\text{ref}(x,y;t) \ar[dr]^{t=-1} & & A^{\text{Q-def}} (x,y;a) \ar[dl]_{a=1} \\
& A(x,y) &
}
\label{AAAA}
\eeq
which should remind the reader of the diagram \eqref{poincare-diagram} expressing a similar relation between various polynomial and homological invariants discussed here. Indeed, each of the invariants in \eqref{poincare-diagram} has a $n$-colored analog, whose color dependence is controlled by the corresponding deformation of the $A$-polynomial in \eqref{AAAA}. In this diagram, we included yet another deformation of the $A$-polynomial, which can be obtained from the super-$A$-polynomial by setting $t=-1$. This so-called \vocab{$Q$-deformation} of the $A$-polynomial was recently studied in \cite{AVqdef}, where it was conjectured that $A^{\text{Q-def}} (x,y;a)$ agrees with the augmentation polynomial
of \vocab{knot contact homology}~\cite{NgFramed,EENS,Ng}.


As a closing remark, we should mention that the colored homological invariants have even more structure than we have so far discussed. One can also construct a family of \vocab{colored differentials}, which act by removing boxes from Young tableaux or reducing the dimension of the representation in the decoration of a link diagram \cite{GS}. For example,
\beq
(\mathscr{H}^{\square\!\square}, d_\text{colored}) \simeq \mathscr{H}^\square,
\eeq
where $
(\mathscr{H}^{\square\!\square}, d_\text{colored})$ denotes the homology of the complex with respect to  the indicated differential.
This can be expressed for the respective Poincar\'e polynomials by a relation of the form \eqref{d1diff}:
\beq
\mathscr{P}~^{\square\!\square}(a,q,t) = a^s \mathscr{P}^\square(a,q^2,t) + (1+at) Q_+(a,q,t),
\eeq
showing the color dependence of these invariants in the form that nicely integrates with the recursion \eqref{superAPconj}.

In general, there are many more colored differentials, which altogether form a very rich and rigid structure~\cite{GS}.
To fully appreciate the beauty and the power of this structure one needs to consider \vocab{homologically thick knots}.
Roughly speaking, these are the knots whose homological invariants contain a lot more new information compared
to their polynomial predecessors.
The knot $8_{19} = T^{(3,4)}$  --- that can be equivalently viewed as a $(3,4)$ torus knot --- is the first example
of a homologically thick knot. Other examples of homologically thick knots and links include \vocab{mutants}.

In the case of $n$-colored HOMFLY homology that we discussed earlier,
the colored differentials include the differentials $d_N$ of section \ref{sec:categorification}
for special values of $N$ in the range $-2n+3, \ldots, 1$.
Note, in the uncolored theory ($n=2$) this range contains only three differentials, $d_{\pm 1}$ and $d_0$,
which play a very special role.
Namely, the first two are canceling differentials, whereas $d_0$ is the differential
that relates HOMFLY homology to \vocab{knot Floer homology}~\cite{DGR}.
We emphasize that the last relation really requires the knowledge of how $d_0$ acts on HOMFLY homology,
which is an extra data not contained in the Poincar\'e polynomial $\mathscr{P} (a,q,t)$.
Curiously, this extra data is automatically contained in the colored version of the HOMFLY homology,
so that \vocab{knot Floer homology} can be recovered directly from $\mathscr{P}_n(a,q,t)$,
even for homologically thick knots!

The reason for this is that all three special differentials $d_1$, $d_{-1}$, and $d_0$,
have analogs in the $n$-colored theory.
Moreover, they are part of the colored differentials $d_N$, with $N = -2n+3, \ldots, 1$.
Specifically, in the $n$-colored HOMFLY homology the differentials $d_1$ and $d_{1-n}$ are canceling,
whereas $d_{2-n}$ provides the relation to knot Floer homology~\cite{GS,inprogress}.
And the virtue of the colored theory is that the action of this latter differential can be
deduced from the data of $\mathscr{P}_n(a,q,t)$ alone.
In other words, what in the uncolored theory appears as a somewhat bizarre and irregular behavior at $N=-1,0,+1$
becomes a natural and simple structure in the colored theory.

\nocite{*}

\printindex

\bibliography{quant-cat-notes}
\bibliographystyle{amsplain}


\end{document}

%% file: preamble.tex
\usepackage{setspace}
\usepackage{mathrsfs}
\usepackage{amssymb}
\usepackage{amsmath}

\usepackage[all,cmtip,knot]{xy}
\entrymodifiers={+!!<0pt,\fontdimen22\textfont2>}

\usepackage{tikz}

\def\ctb{T^*\!} 		

\def\beq{\begin{equation}}
\def\eeq{\end{equation}}

\def\lie{\mathcal{L}}		

\def\Cinf{C^\infty}
\def\R{\mathbb{R}}
\def\C{\mathbb{C}}
\def\Z{\mathbb{Z}}
\def\Q{\mathbb{Q}}

\def\im{\qopname \relax o{im}}
\def\inj{\hookrightarrow}

\def\goesto{\rightarrow}
\def\isomto
{\mathrel{\tilde{\rightarrow}}}
\def\isom{\mathrel{\widetilde{=}}}

\DeclareMathOperator{\Tr}{Tr} 		
\DeclareMathOperator{\Hom}{Hom}

\def\GL(#1,#2){{\mathrm{GL}_{#1}#2}}
\def\SL(#1,#2){{\mathrm{SL}_{#1}#2}}
\def\Sp(#1){\mathrm{Sp}(#1)}
\def\sp(#1){\mathfrak{sp}(#1)}	

\def\SU{\mathrm{SU}}

\def\defeq{\mathrel{:=}}

\theoremstyle{definition}
\newtheorem{example}{Example}
\newtheorem{exercise}{Exercise}
\newtheorem{remark}{Remark}
\newtheorem{theorem}{Theorem}

\usepackage{graphicx,graphics}
\usetikzlibrary{intersections}

\def\Surf{\Sigma}
\def\lie{\mathfrak}
\DeclareMathOperator{\rank}{rank}
\def\bdy{\partial}
\def\op{\hat}
\def\E{\mathscr{E}}
\def\Hil{\mathscr{H}}
\def\moduli{\mathscr{M}}
\def\Khom{H_{i,j}}
\DeclareMathOperator{\Kh}{Kh}
\def\rep#1{\mathbf{#1}}
